\documentclass{appolb}
\usepackage{epsfig}
\newcommand{\lesssim}{\raisebox{0.3mm}{\em $\, <$} 
\hspace{-3.3mm} \raisebox{-1.8mm}{\em $\sim \,$}}
\newcommand{\gtrsim}{\raisebox{0.3mm}{\em $\, >$}
\hspace{-3.3mm} \raisebox{-1.8mm}{\em $\sim \,$}}
\begin{document}
\pagestyle{plain}

\title{Three Flavor Neutrino Oscillations and\\
Application to Long Baseline Experiments
\thanks{Invited talk at the XXIII International School of Theoretical
Physics, September 15--21, Ustron, Poland.
}
}
\author{Osamu Yasuda\footnote{Email: yasuda@phys.metro-u.ac.jp}
\address{Department of Physics,
Tokyo Metropolitan University \\
Minami-Osawa, Hachioji, Tokyo 192-0397, Japan}
}
\maketitle

\begin{abstract}
Using the result of the three flavor analysis of the old
Kamiokande data, the recent
Superkamiokande data of atmospheric neutrinos and
the CHOOZ reactor data, it is shown that
the third mixing angle $\theta_{13}$ is small.
It is proposed to determine the small value of $\theta_{13}$
and the CP violating phase $\delta$
in very long baseline experiments by measuring the
appearance probability $P(\nu_\mu\rightarrow\nu_e)$
and the T violating effect
$P(\nu_e\rightarrow\nu_\mu)-P(\nu_\mu\rightarrow\nu_e)$ which
are enhanced by the matter effect of the Earth.
\end{abstract}
\vskip 0.5cm

\section{Introduction}
The solar neutrino data \cite{homestake,Kamsol,SKsol,sage,gallex} and
the atmospheric neutrino experiments
\cite{Kamatm,IMB,SKatm,SKup,soudan2} provide strong evidence for
neutrino oscillations.  In the framework of the two flavor neutrinos,
these experimental data are explained by two sets of the oscillation
parameters $(\Delta m^2_\odot,\sin^22\theta_\odot)\simeq$ $({\cal
O}(10^{-5}{\rm eV}^2),{\cal O}(10^{-2}))$ (small angle MSW solution),
$({\cal O}(10^{-5}{\rm eV}^2),{\cal O}(1))$ (large angle MSW
solution), or $({\cal O}(10^{-10}{\rm eV}^2),{\cal O}(1))$ (vacuum
oscillation solution), and $(\Delta m_{\mbox{\rm
atm}}^2,~\sin^22\theta_{\mbox{\rm atm}})\simeq (10^{-2.5}{\rm
eV}^2,1.0)$.

\begin{figure}
\vglue -1.0cm
\hglue 3.0cm\epsfig{file=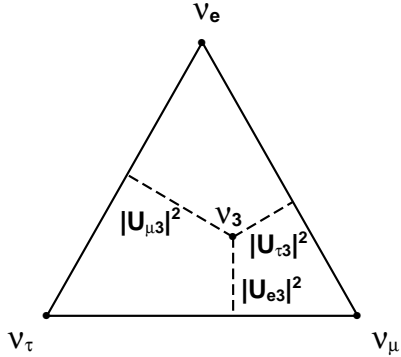,width=6cm}
\vglue -1cm
\caption{Triangle of unit height for $\nu_3$.
The vertical position of the state in the triangle
graph is $|U_{e3}|^2$, which indicates deviation from the two flavor
mixings.}
\label{fig:tri}
\end{figure}
Without loss of generality we assume that
$|\Delta m_{21}^2|<|\Delta m_{32}^2|<|\Delta m_{31}^2|$
where $\Delta m^2_{ij}\equiv m^2_i-m^2_j$.
If both the solar neutrino deficit and the atmospheric neutrino 
anomaly are to be solved by energy dependent solutions, we have to have
$\Delta m_{21}^2\simeq\Delta m^2_\odot$ and
$\Delta m_{32}^2\simeq\Delta m^2_{\mbox{\rm atm}}$, i.e.,
we have mass hierarchy in this case.
Therefore I will assume mass hierarchy in the
three flavor framework throughout this talk.
I will adopt the triangle representation which has been introduced by
Fogli, Lisi, and Scioscia \cite{FLS}.
Fig. \ref{fig:tri}, which will play a role
in the analysis of atmospheric neutrino data, represents
how the most massive state $\nu_3$ mixes with
three flavor eigenstates with the coefficients
$|U_{\alpha 3}|^2$ ($\alpha=e,\mu,\tau$),
where $|U_{\alpha 3}|^2$ are the elements
of the MNS mixing matrix U \cite{mns}:
\begin{eqnarray}
\left( \begin{array}{c} \nu_e  \\ \nu_{\mu} \\ 
\nu_{\tau} \end{array} \right)
=U\left( \begin{array}{c} \nu_1  \\ \nu_2 \\ 
\nu_3 \end{array} \right),\quad
U&\equiv&\left(
\begin{array}{ccc}
U_{e1} & U_{e2} &  U_{e3}\\
U_{\mu 1} & U_{\mu 2} & U_{\mu 3} \\
U_{\tau 1} & U_{\tau 2} & U_{\tau 3}
\end{array}\right).
\label{eqn:mns}
\end{eqnarray}

\section{Constraints from atmospheric neutrino anomaly}
\begin{figure}
\vglue 0cm \hglue -1cm \epsfig{file=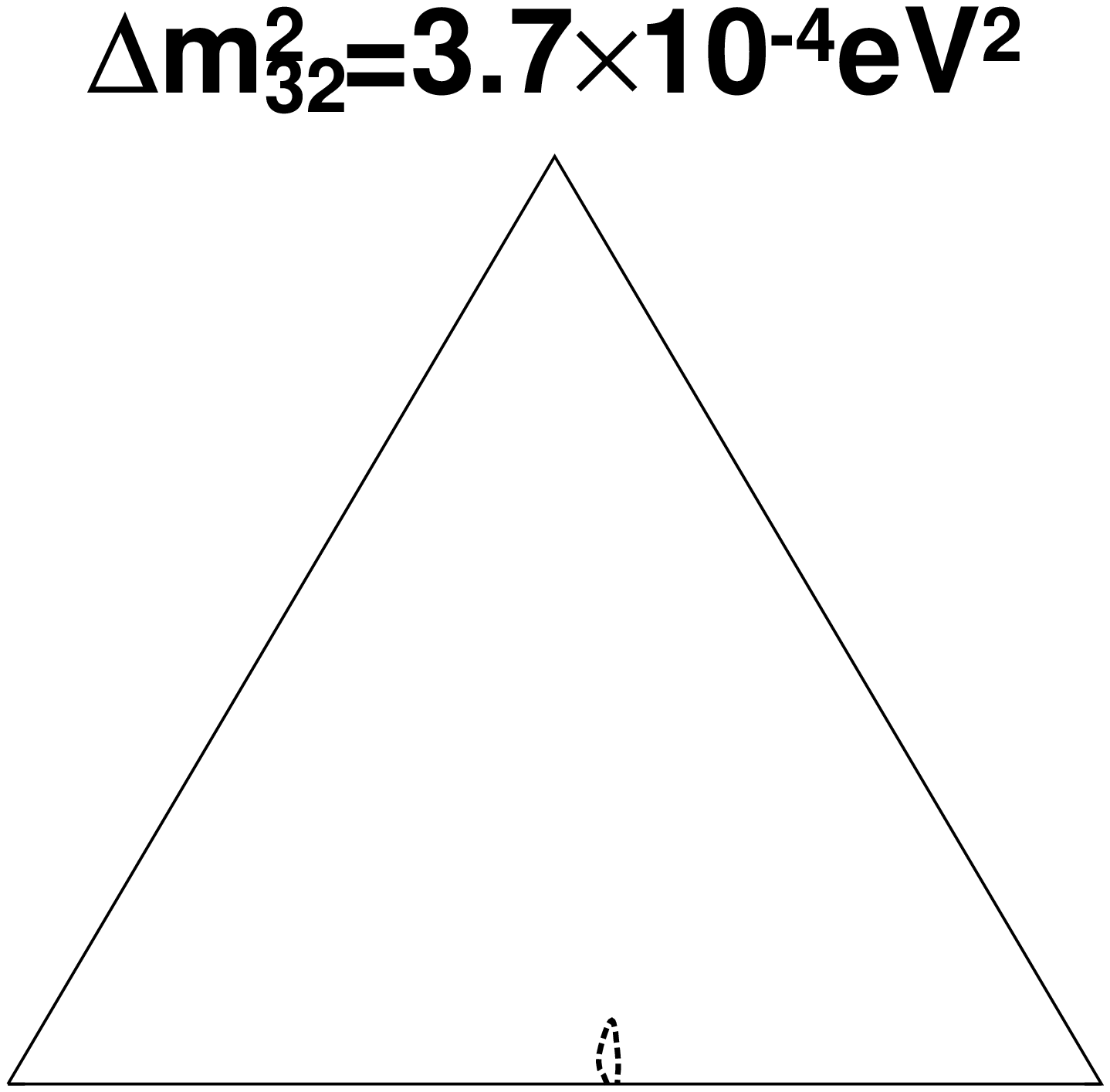,width=5cm}
\vglue -4.65cm \hglue 3.8cm \epsfig{file=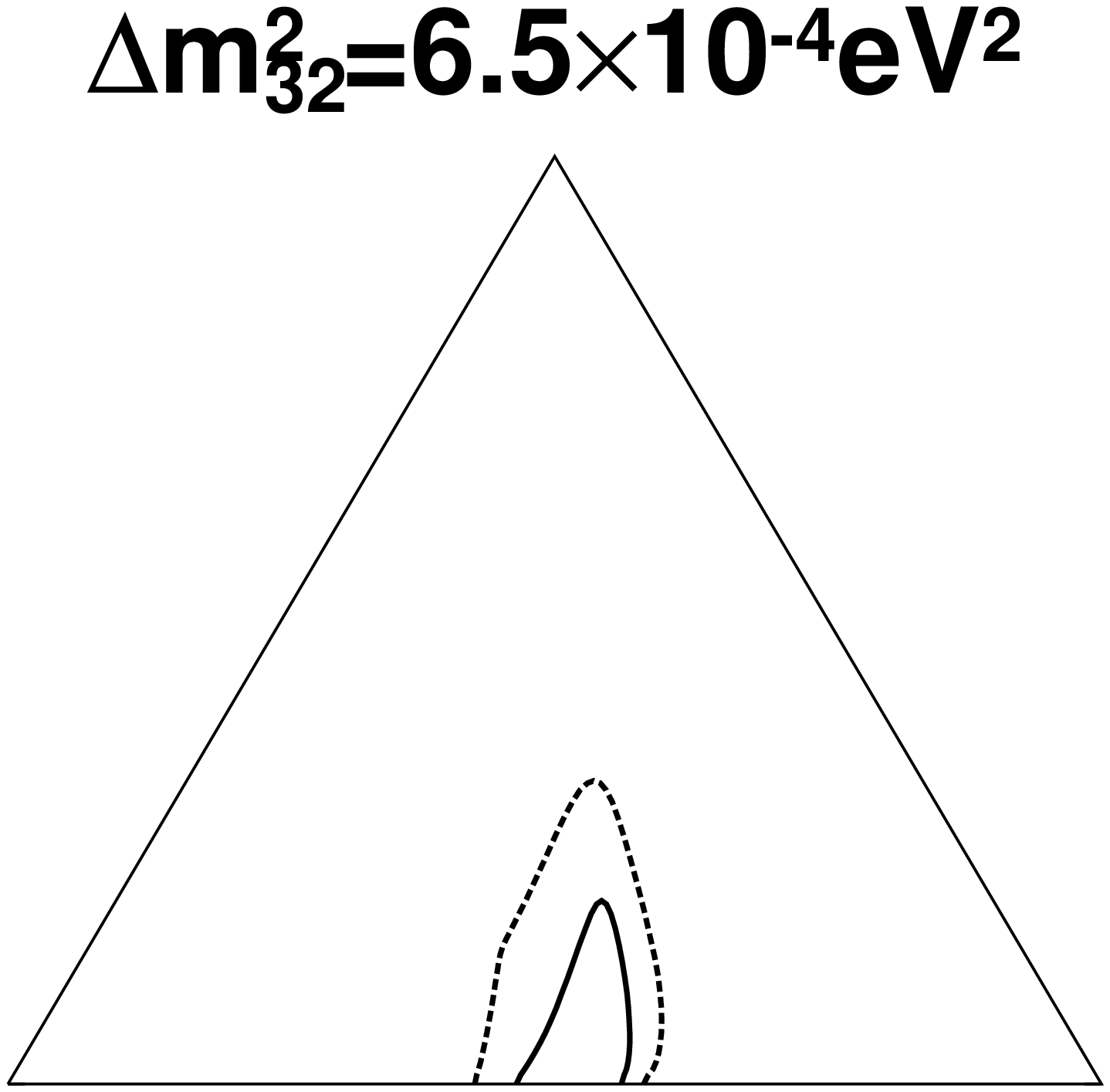,width=5cm}
\vglue -4.65cm \hglue 8.8cm \epsfig{file=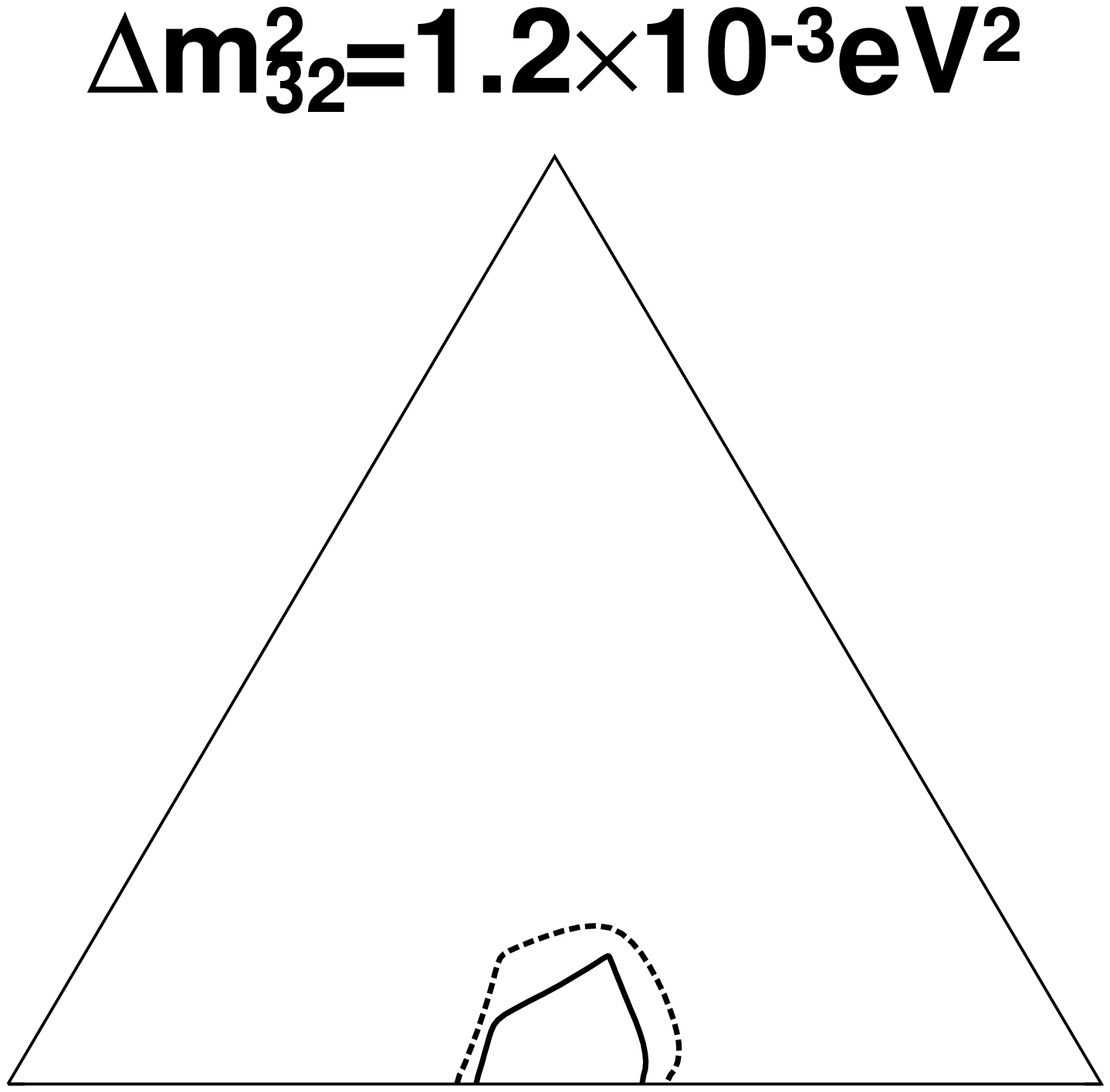,width=5cm}
\vglue 0cm\hglue -1cm \epsfig{file=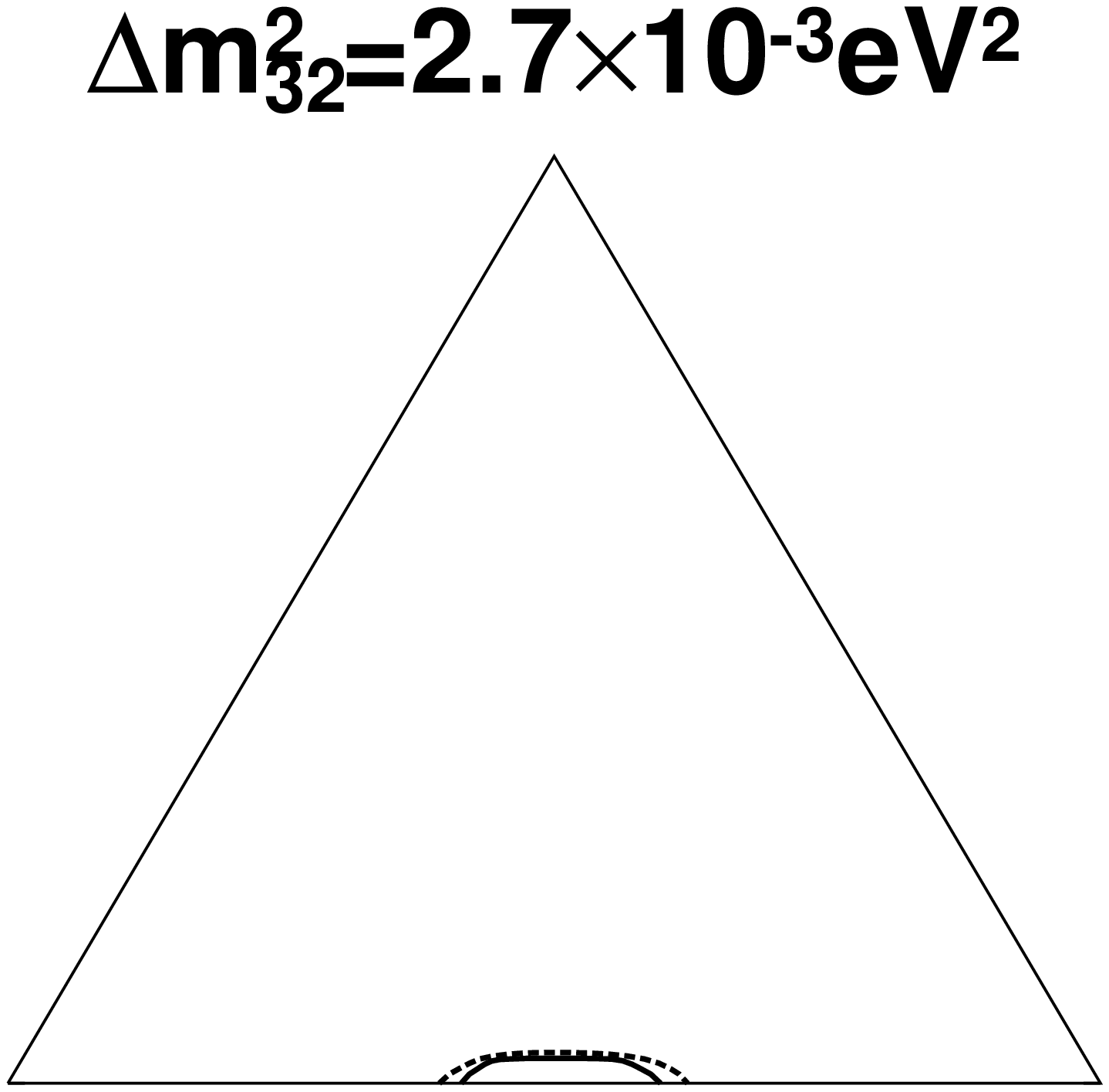,width=5cm}
\vglue -4.65cm \hglue 3.8cm \epsfig{file=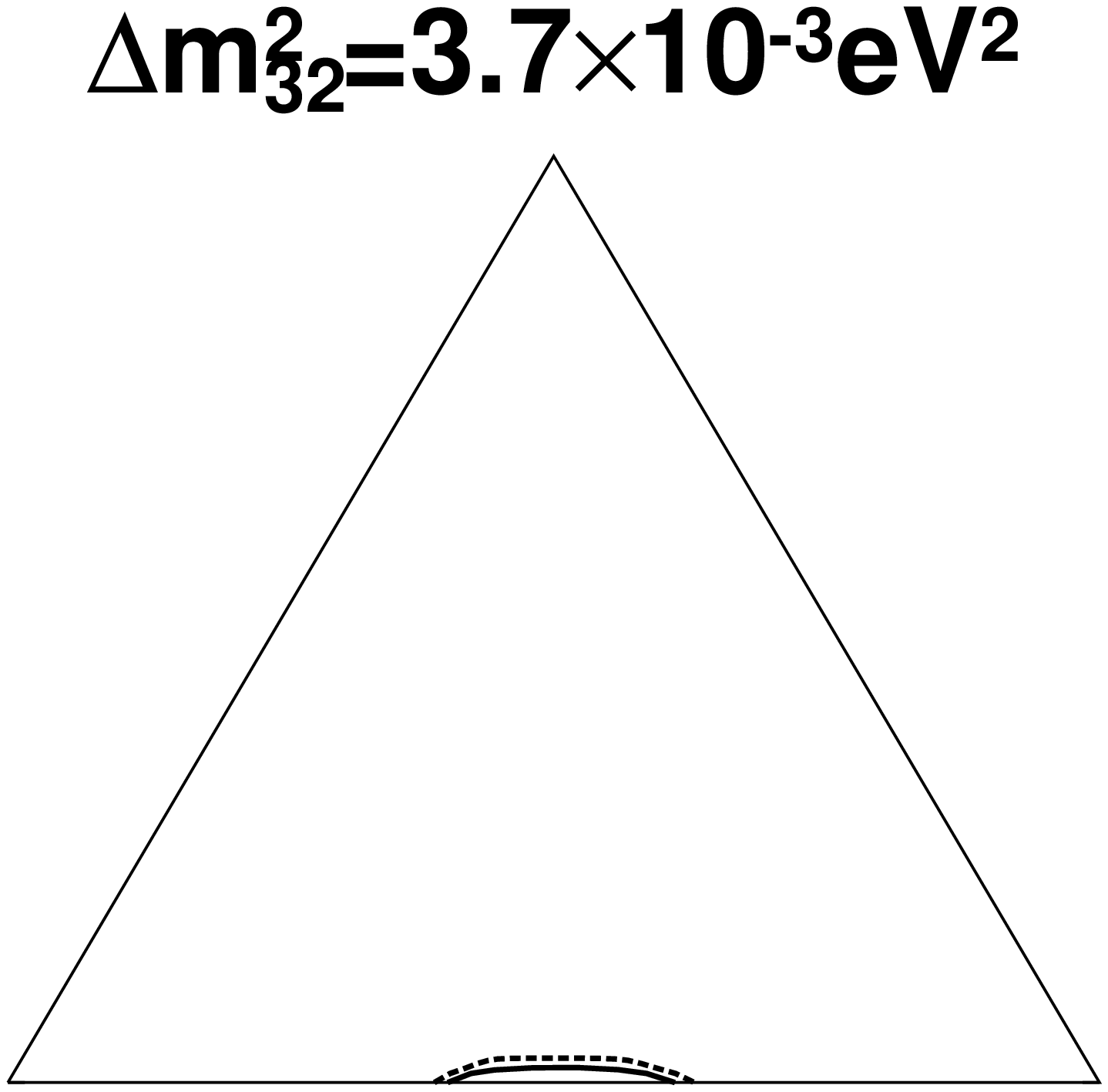,width=5cm}
\vglue -4.65cm \hglue 8.8cm \epsfig{file=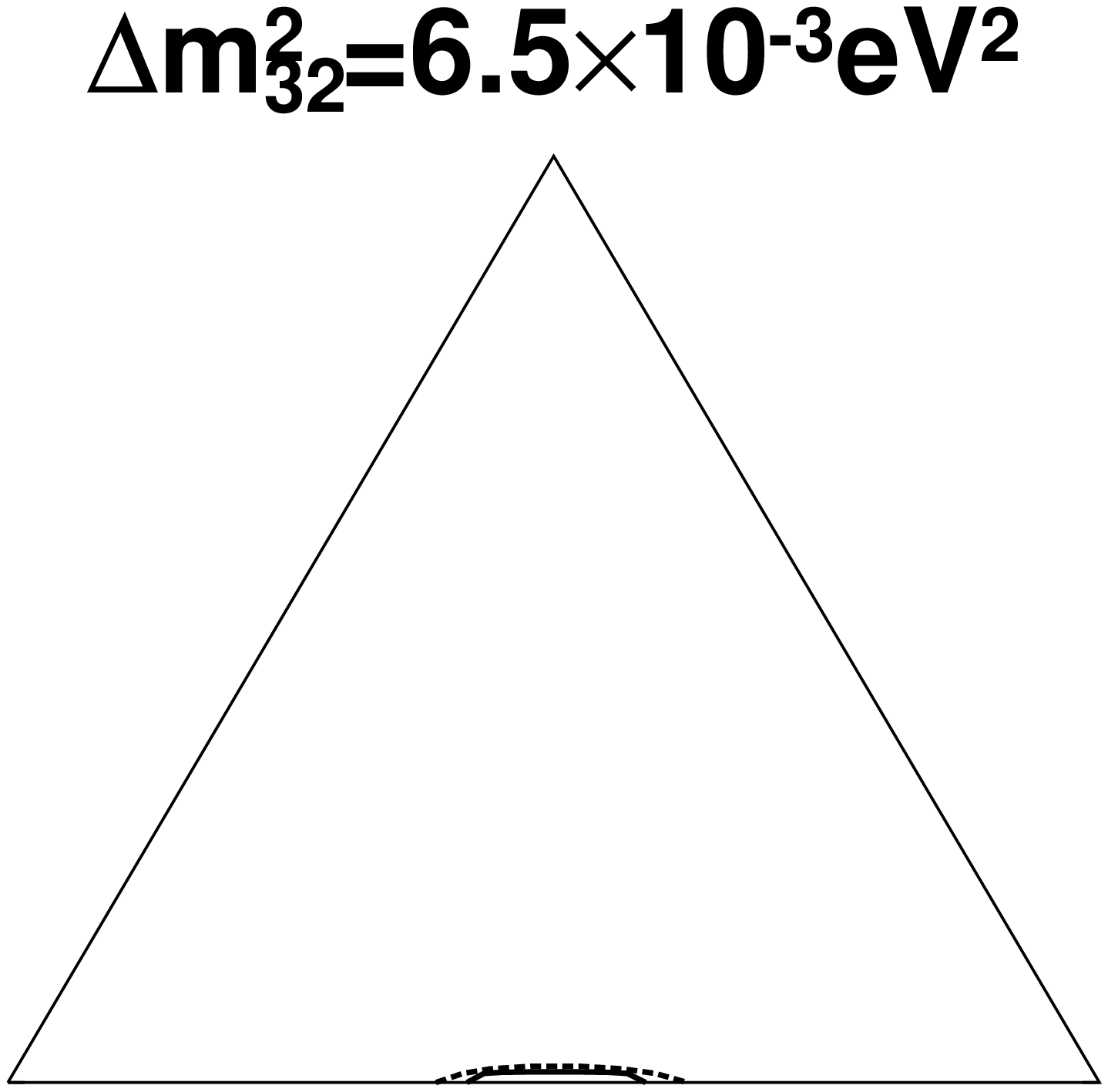,width=5cm}
\vglue 0cm \hglue -1cm \epsfig{file=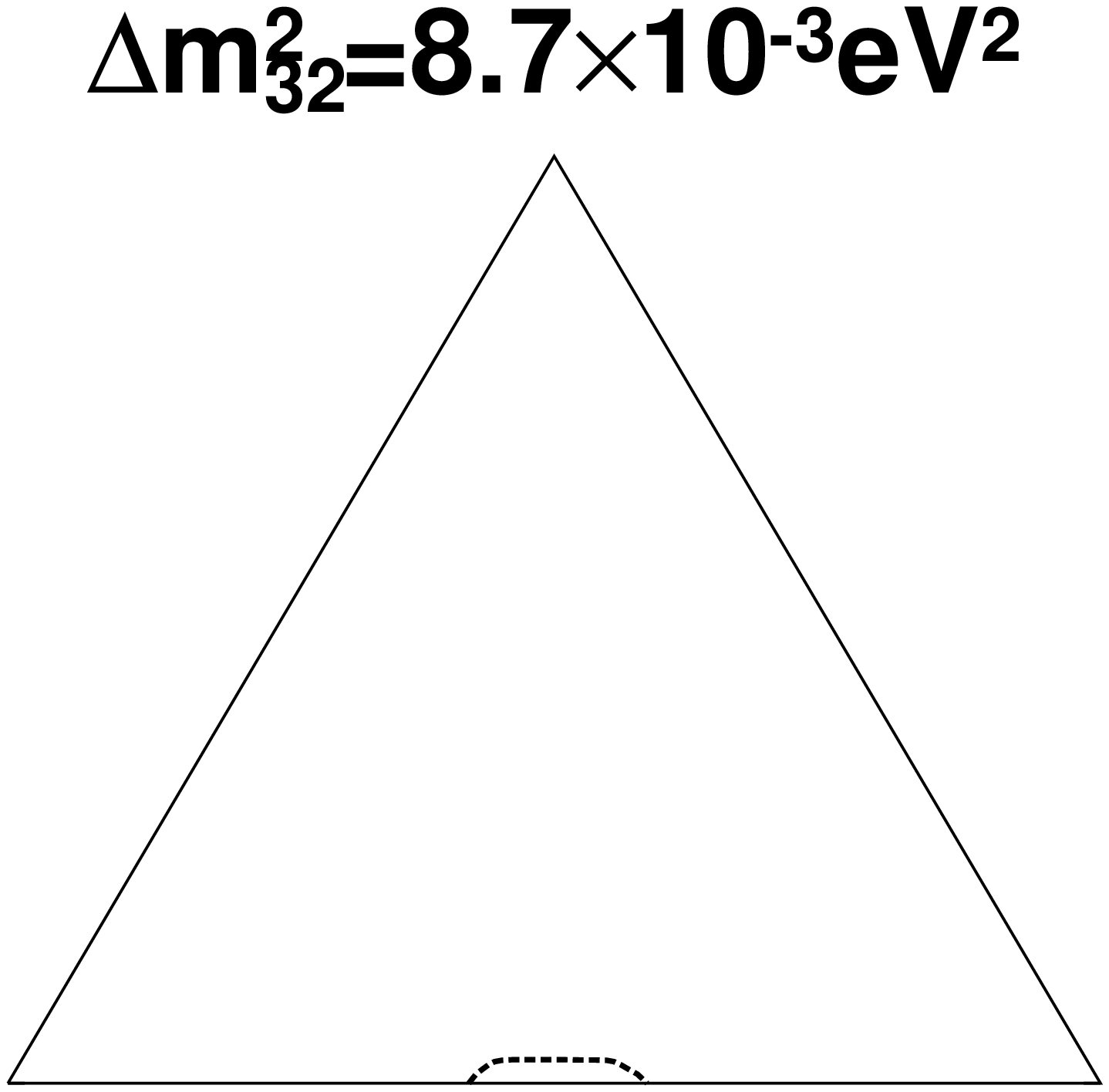,width=5cm}
\vglue -4.65cm \hglue 3.8cm \epsfig{file=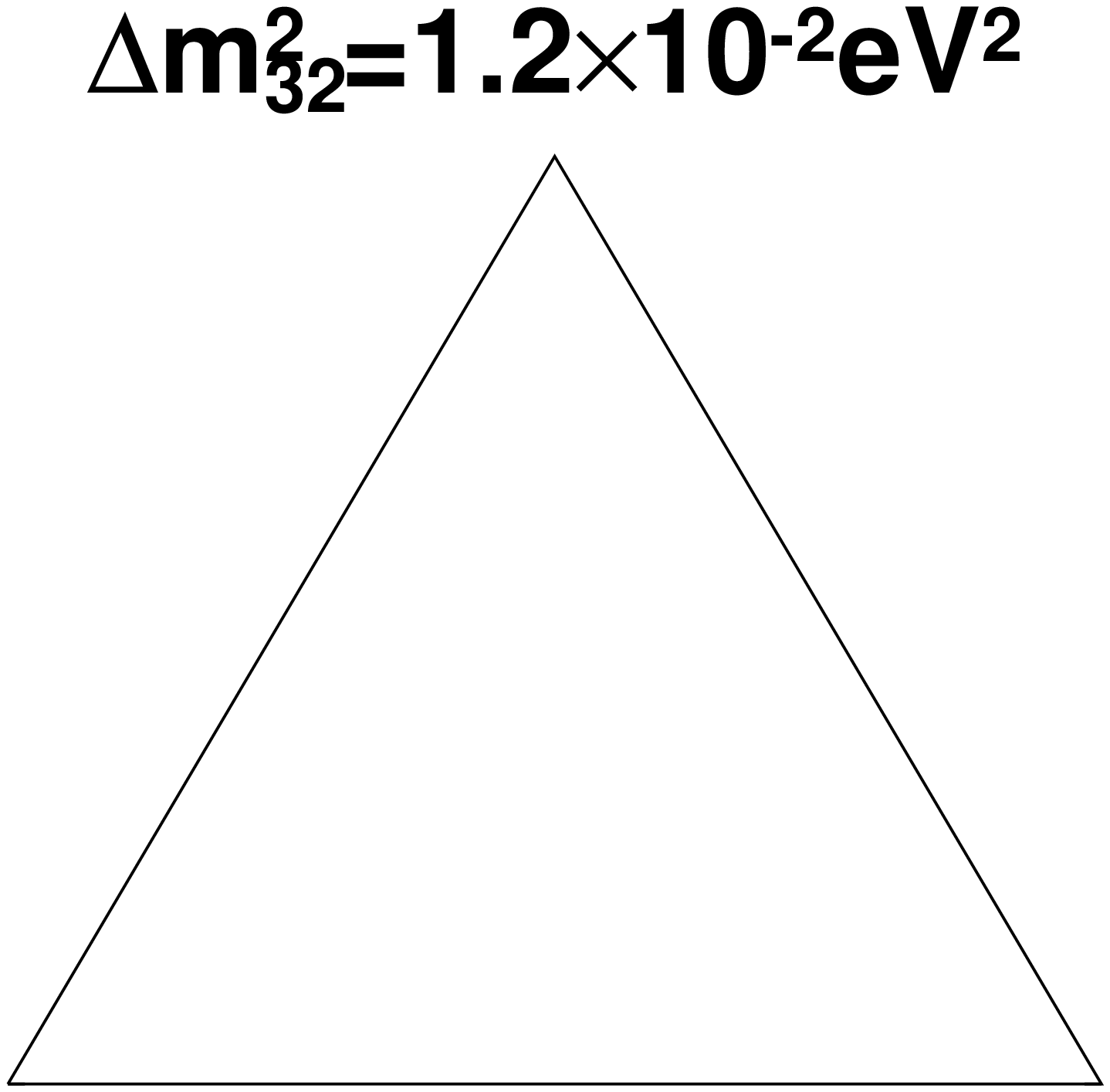,width=5cm}
\vglue -4.65cm \hglue 8.8cm \epsfig{file=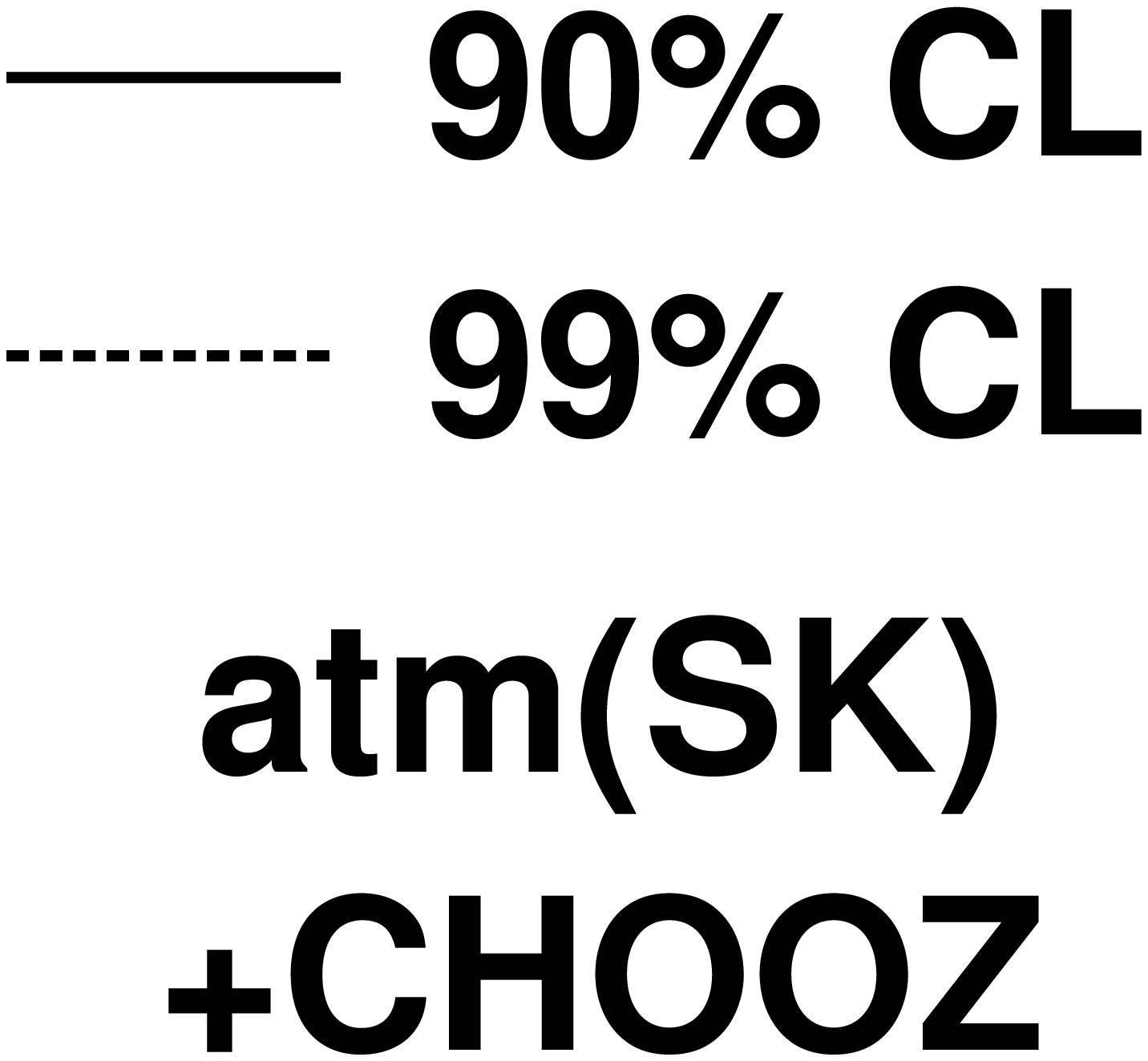,width=5cm}
\caption{The allowed regions
for various $\Delta m^2_{32}$ by the
constraints of atmospheric neutrino data of
the Superkamiokande
contained and upward going $\mu$ events, and
the CHOOZ reactor data.}
\label{fig:skatm}
\end{figure}
\begin{figure}
\vglue 0cm \hglue -1cm \epsfig{file=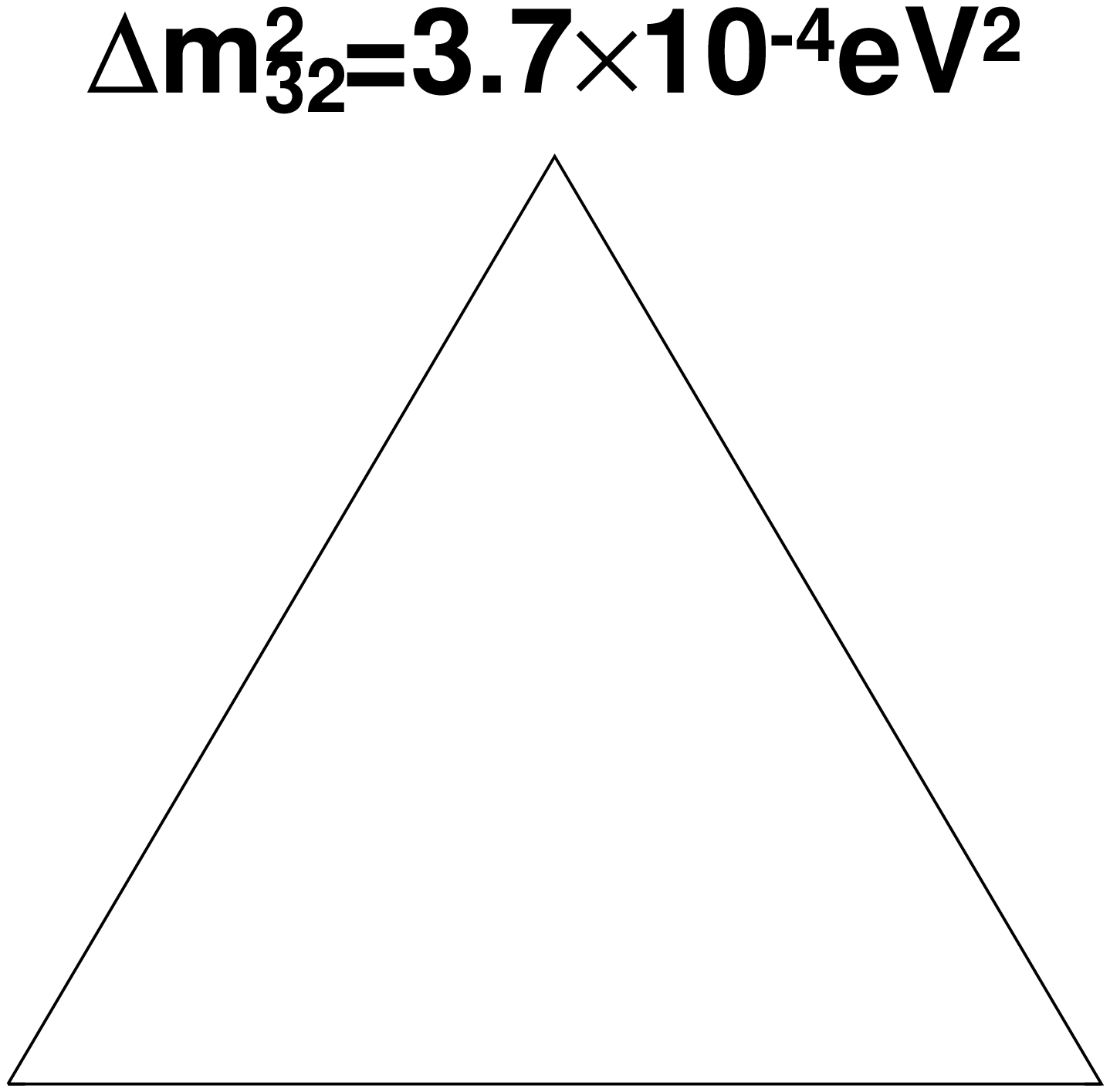,width=5cm}
\vglue -4.65cm \hglue 3.8cm \epsfig{file=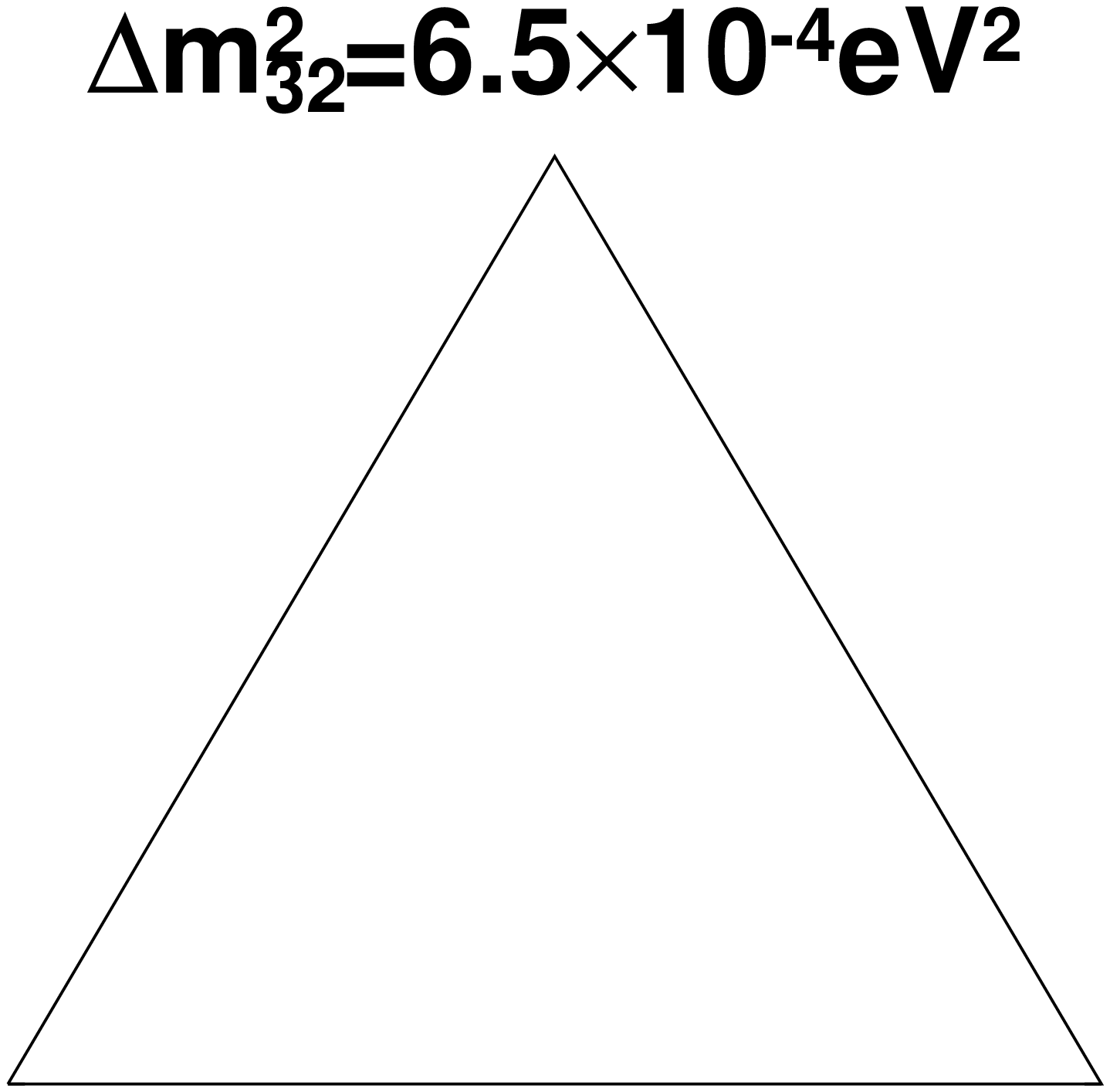,width=5cm}
\vglue -4.65cm \hglue 8.8cm \epsfig{file=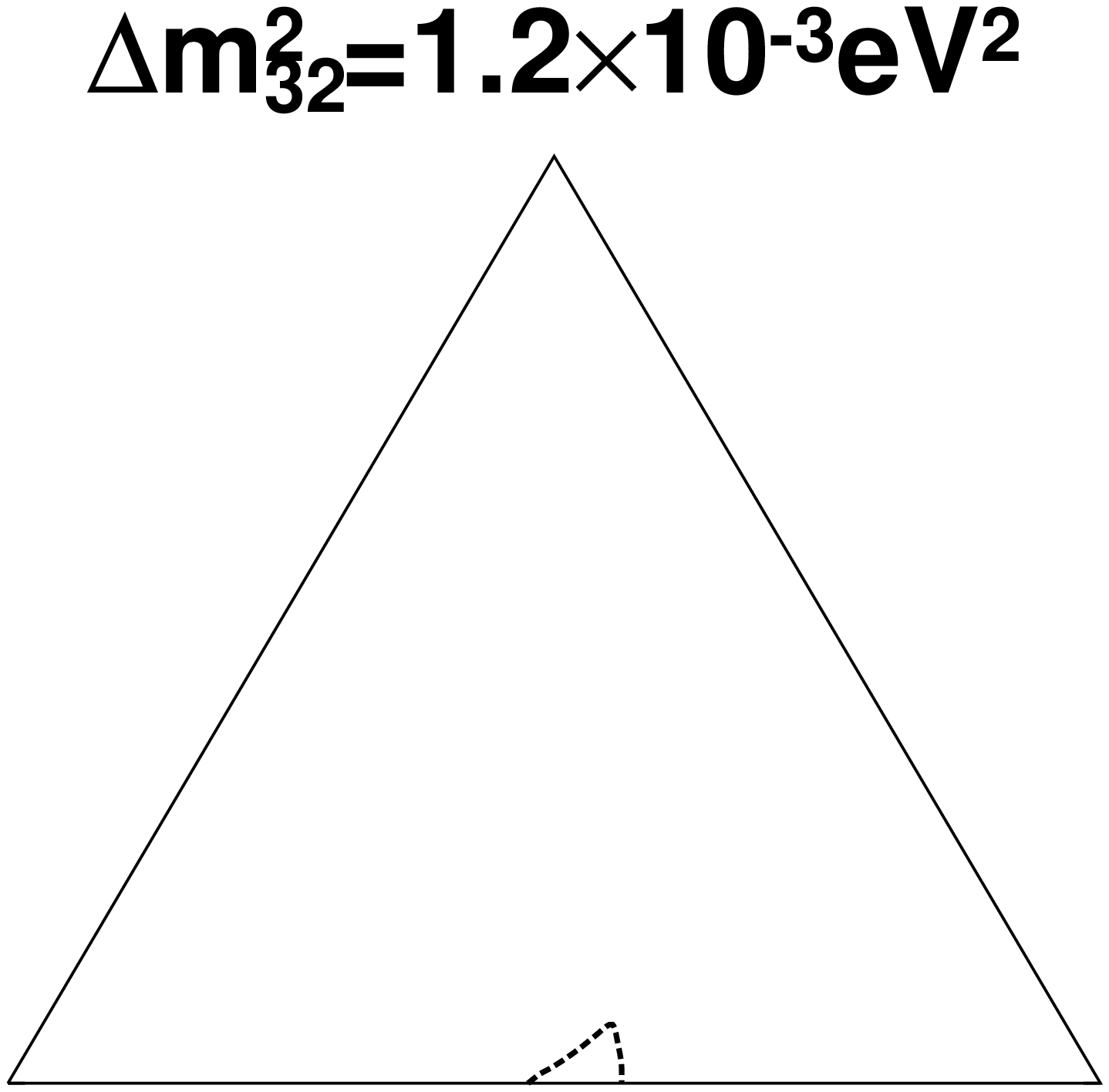,width=5cm}
\vglue 0cm \hglue -1cm \epsfig{file=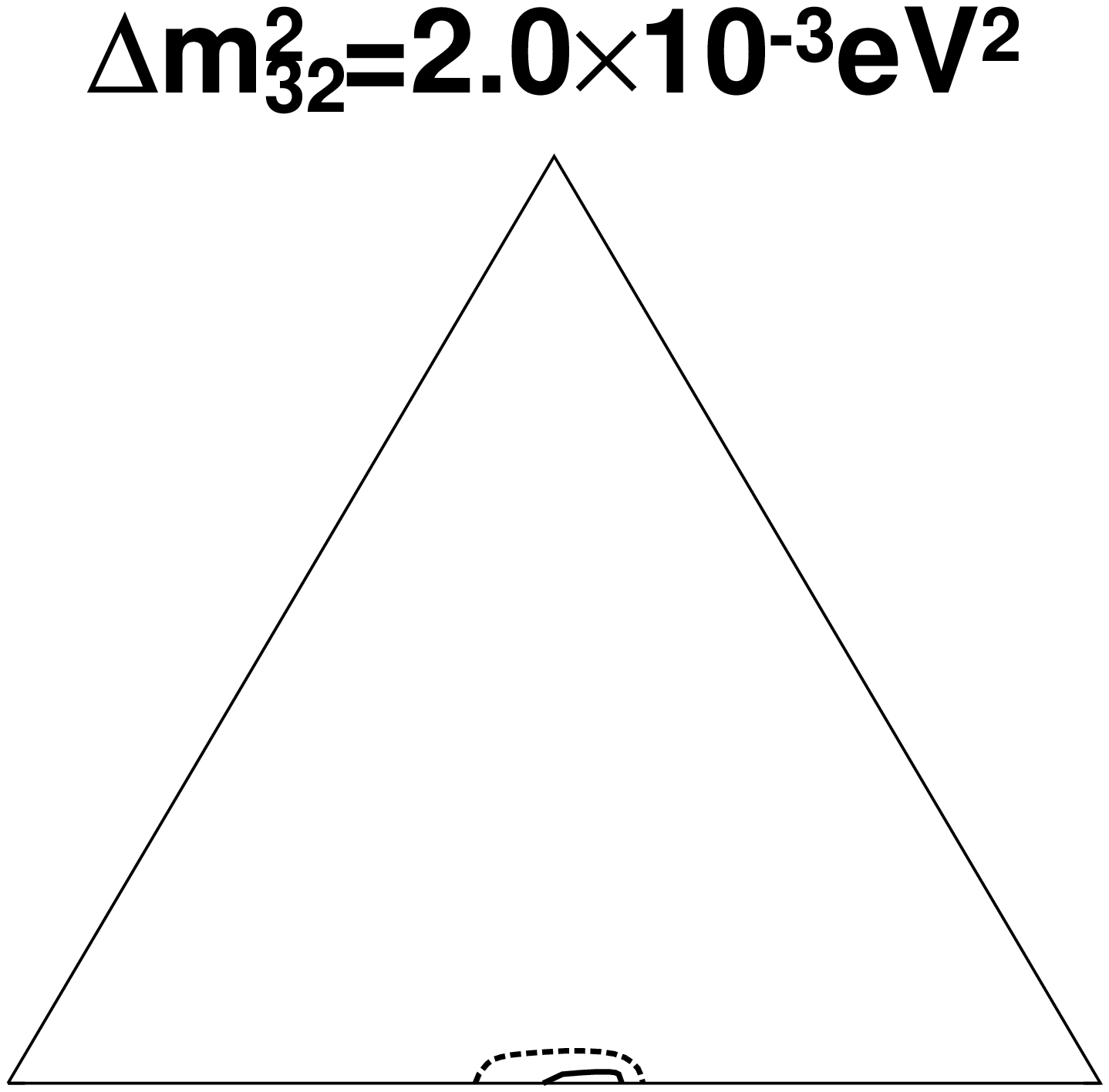,width=5cm}
\vglue -4.65cm \hglue 3.8cm \epsfig{file=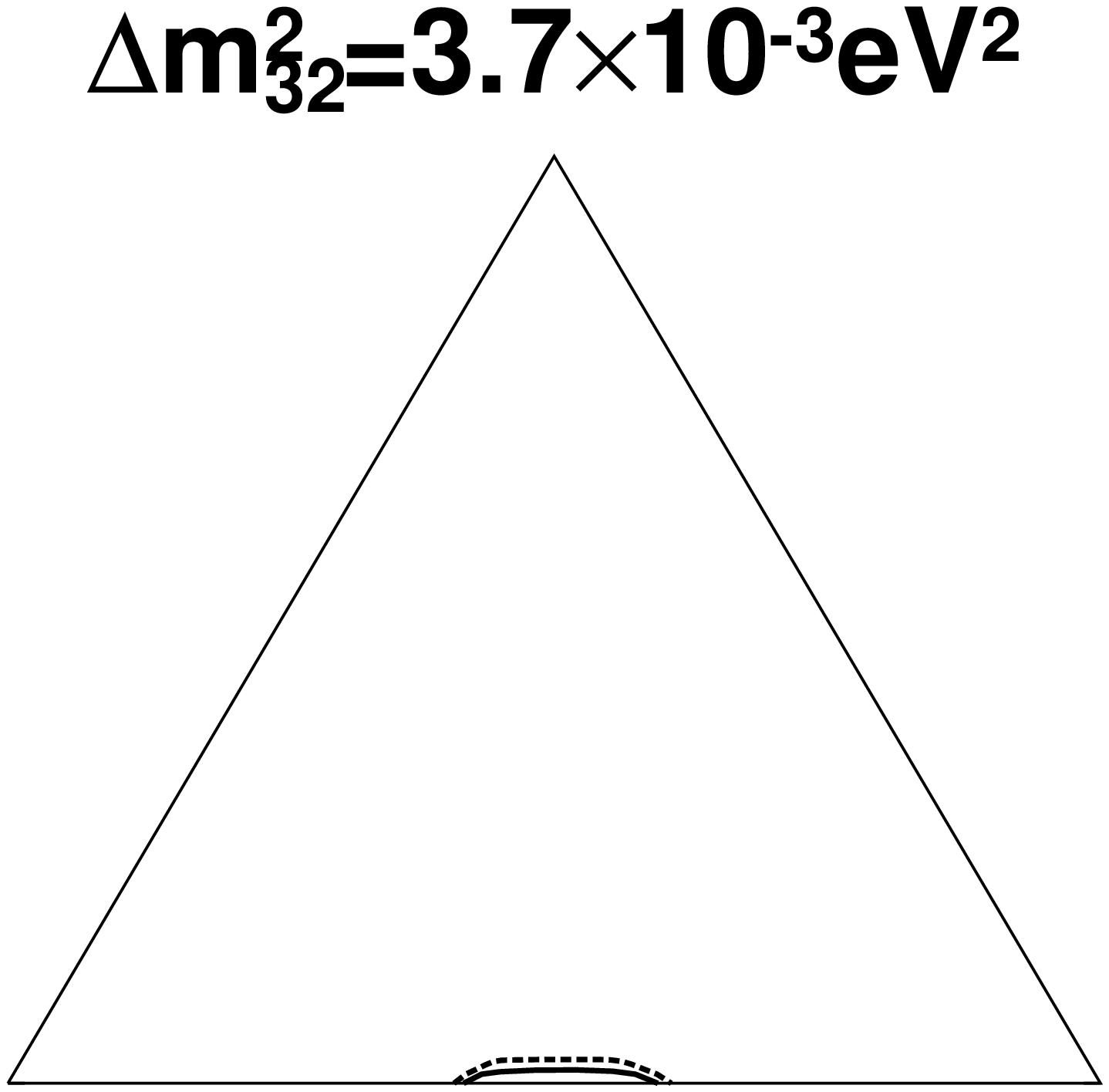,width=5cm}
\vglue -4.65cm \hglue 8.8cm \epsfig{file=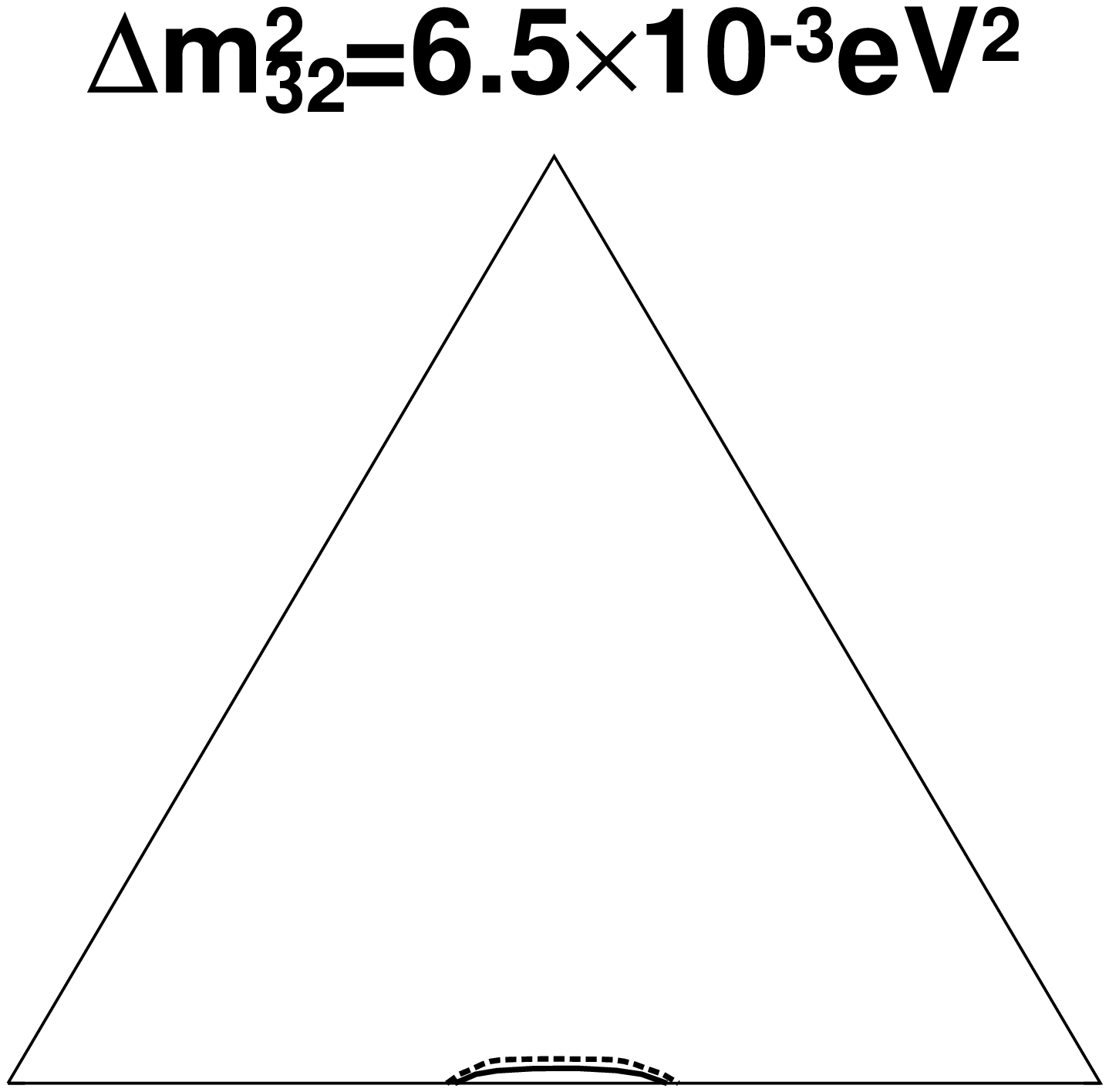,width=5cm}
\vglue 0cm \hglue -1cm \epsfig{file=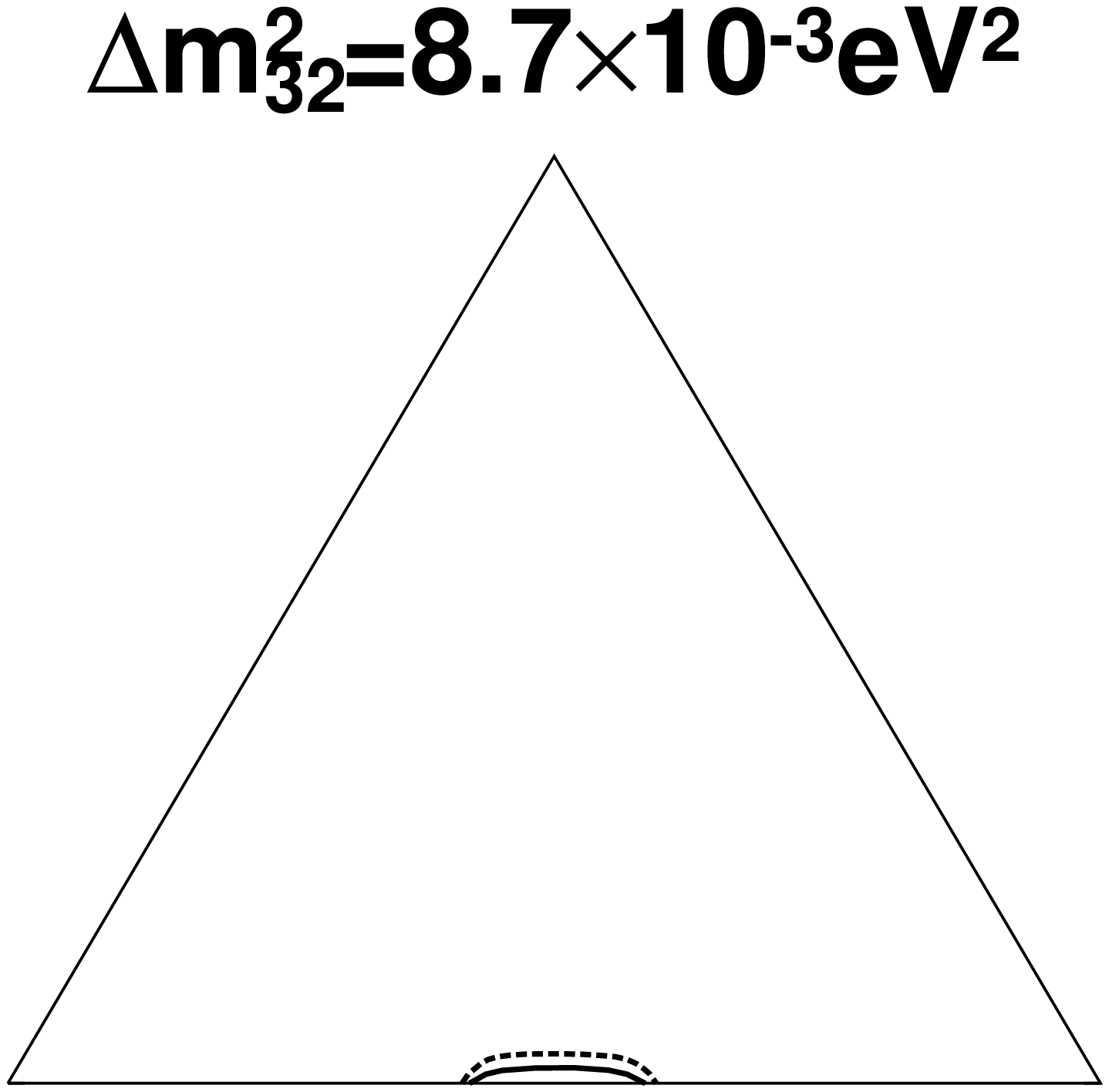,width=5cm}
\vglue -4.65cm \hglue 3.8cm \epsfig{file=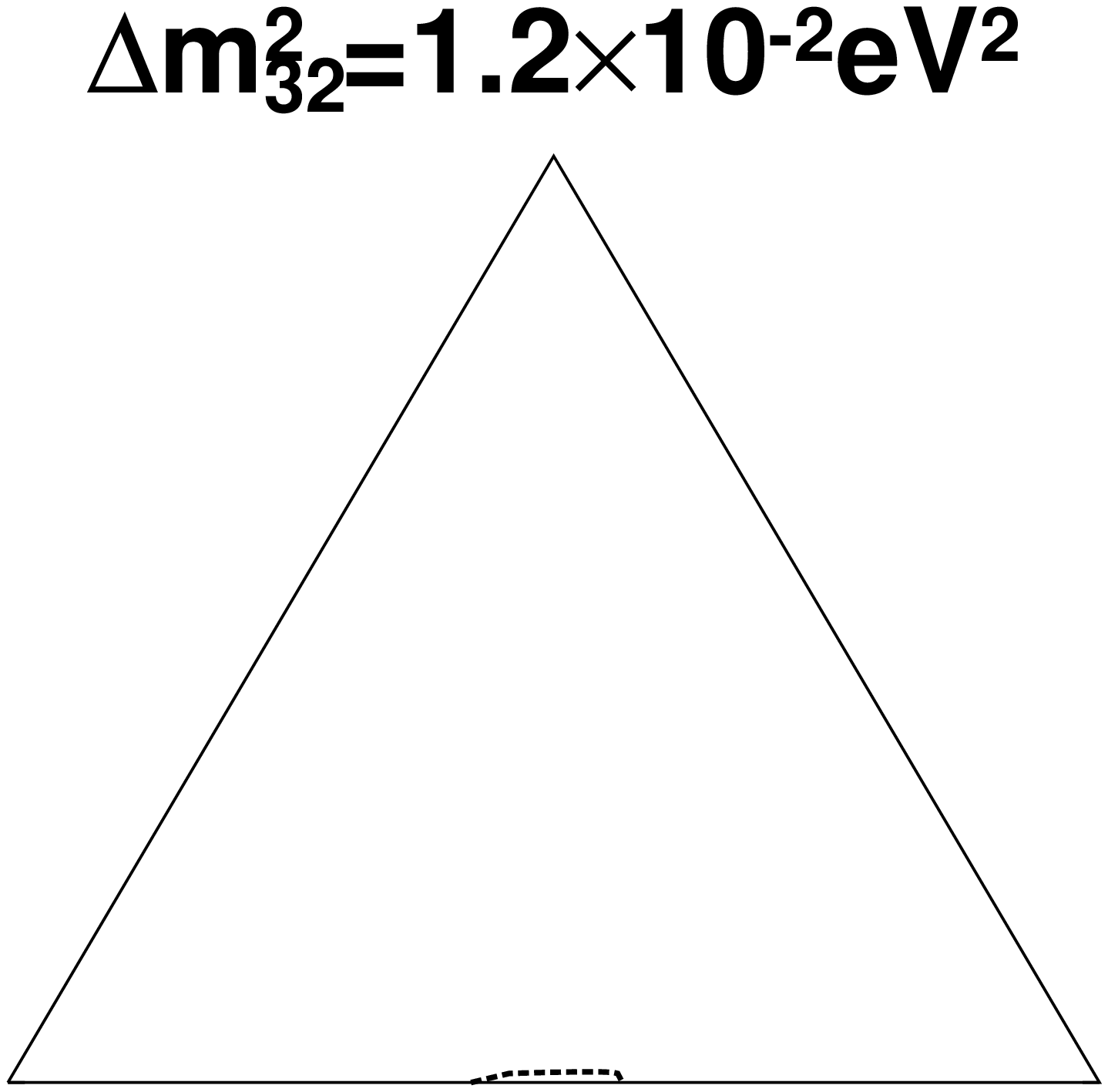,width=5cm}
\vglue -4.65cm \hglue 8.8cm \epsfig{file=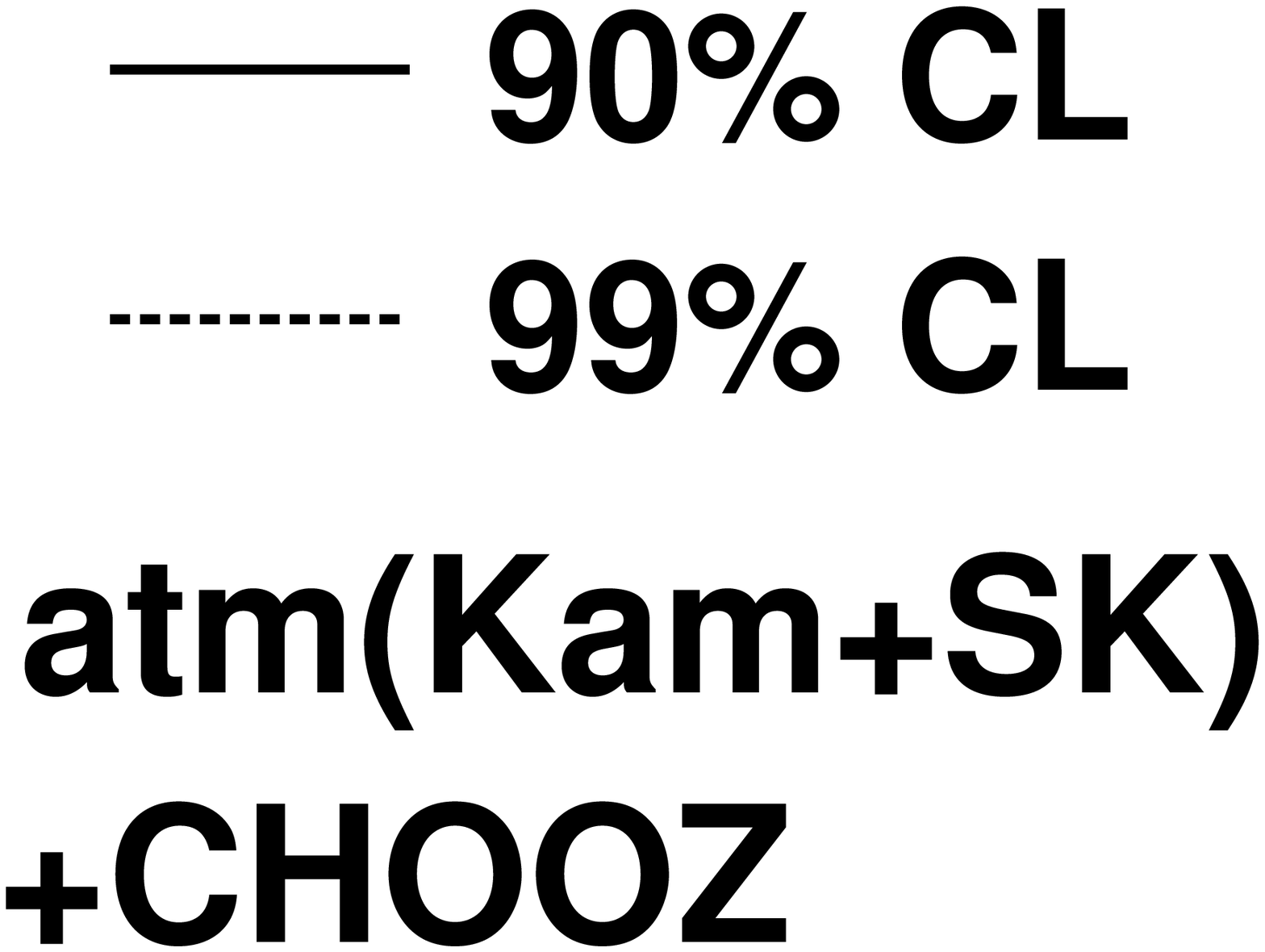,width=5cm}
\caption{The allowed regions
for various $\Delta m^2_{32}$ by the
constraints of atmospheric neutrino data of
the Kamiokande contained events, the Superkamiokande
contained and upward going $\mu$ events, and
the CHOOZ reactor data.  All the shadowed regions are located
near the $\nu_\mu-\nu_\tau$ line.}
\label{fig:kamatm}
\end{figure}

The atmospheric neutrino data of Superkamiokande have been analyzed by
\cite{y1,FLMS} in the three flavor framework, where smaller mass
squared difference $\Delta m_{21}^2$ is ignored.  The two flavor
analysis of the most up-to-date data by the Superkamiokande group
shows that the allowed region of the mass squared difference is
$1\times 10^{-3}$eV$^2<\Delta m^2 < 7\times 10^{-3}$eV$^2$ at 90\% CL
\cite{SKatm}.  The analyses in \cite{y1,FLMS} are strictly speaking
different from the original one in \cite{SKatm}, since the full data
which are binned with respect to the energy as well as the zenith
angle are not used in \cite{y1,FLMS}.  The analysis in \cite{y1} has
been updated with the recent data for 850 days, where the upward going
$\mu$ data \cite{SKup} have also been incorporated.  It was found that
the region of the mass squared difference which is as small as
$5\times 10^{-4}$eV$^2$ is allowed at 90\%CL
(cf. Fig. \ref{fig:skatm}).

On the other hand, the CHOOZ group has updated their result on
$P(\bar{\nu}_e\rightarrow\bar{\nu}_e)$ in the reactor disappearance
experiment \cite{chooz}, and the mass squared difference is limited to
$\Delta m^2 < 7\times 10^{-4}$eV$^2$ for the maximum mixing.  In our
three flavor scheme with mass hierarchy, the disappearance probability
for the CHOOZ experiment is given by
\begin{eqnarray}
P(\bar{\nu}_e\rightarrow\bar{\nu}_e)=1-\sin^22\theta_{13}\sin^2
\left({\Delta m_{32}^2L \over 4E}\right),
\end{eqnarray}

\hglue -0.6cm
so if $\Delta m_{32}^2>7\times 10^{-4}$eV$^2$ then
$\sin^22\theta_{13}$ has to be small.
The allowed region which is obtained by the constraints of
the Superkamiokande atmospheric neutrino data and
the CHOOZ data is given in Fig. \ref{fig:skatm}.
As is seen in Fig. \ref{fig:skatm}, relatively large
$\theta_{13}$ is still allowed for $\Delta m_{32}^2<1\times 10^{-3}$eV$^2$.

To obtain more stringent bound on $\sin^22\theta_{13}$, I include the
three flavor analysis \cite{y2} of the contained events of the
Kamiokande atmospheric neutrino data \cite{Kamatm}, for which the
value of the mass squared difference in the allowed region tends to be
higher than that of Superkamiokande.  In fact $\Delta m_{32}^2<2\times
10^{-3}$eV$^2$ is excluded at 90\%CL by including the Kamiokande data.
Combining the atmospheric neutrino data of Superkamiokande, Kamiokande
and the CHOOZ reactor data, I have obtained the allowed region which is
depicted in Fig. \ref{fig:kamatm}.  Fig. \ref{fig:kamatm} shows that
$\sin^22\theta_{13} < 0.1$ has to be satisfied, which is basically the
consequence from the CHOOZ data.

\section{A possible way to measure $\theta_{13}$}
As we have seen in sect. 2, the data of
atmospheric neutrinos and the CHOOZ experiment gives
$\sin^22\theta_{13}\lesssim 0.1$.
The two sets of parameters
$(\Delta m^2_{21},\sin^22\theta_{12})$ and
$(\Delta m^2_{32},\sin^22\theta_{23})$ will be determined
with more and more accuracy in the future by various
experiments of solar and atmospheric neutrinos, respectively,
so the next thing we would like to pursue is to determine $\theta_{13}$.
There have been discussions on the future intense muon beam \cite{geer}
which could be hundreds times as high as the present one, and it would
enable us to have very long baseline experiments, where the neutrino
path length is comparable to the radius of the Earth.  In this talk
I would like to point out that the oscillation probability
$P(\nu_\mu\rightarrow\nu_e)$ (in the case of $\Delta m_{32}^2>0$) or
$P({\bar \nu}_\mu\rightarrow{\bar \nu}_e)$ (in the case of $\Delta
m_{32}^2<0$) is enhanced for a certain region of the neutrino energy
due to the matter effect of the Earth when $\sin^22\theta_{13}\gtrsim
0.01$ and therefore it is possible to deduce the magnitude of
$\theta_{13}$ by measuring experimentally $P(\nu_\mu\rightarrow\nu_e)$
or $P({\bar \nu}_\mu\rightarrow{\bar \nu}_e)$ as a function of the
neutrino energy in very long baseline experiments which may be
possible with the intense muon beam technology in the future.

Let us now consider the situation where
$\Delta E_{21}$ is completely negligible.
In that case the positive energy part of the Dirac equation
for three flavors of neutrinos in matter is given by
\begin{eqnarray}
&{\ }&i {d \over dx} \left( \begin{array}{c} \nu_e  \\ \nu_{\mu} \\ 
\nu_{\tau} \end{array} \right)= M
\left( \begin{array}{c} \nu_e  \\
\nu_{\mu} \\ \nu_{\tau}
\end{array} \right)
\end{eqnarray}

\hglue -0.6cm
with
\begin{eqnarray}
M&\equiv&
U \mbox{\rm diag} \left(0,0,\Delta E_{32} \right) U^{-1}
+\mbox{\rm diag} \left(A,0,0 \right)\nonumber\\
&=&D e^{i\theta_{23}\lambda_7}
\left[ e^{i\theta_{13}\lambda_5}
 \mbox{\rm diag} \left(0,0,\Delta E_{32} \right)
e^{-i\theta_{13}\lambda_5}
+\mbox{\rm diag} \left(A,0,0 \right) \right]
e^{-i\theta_{23}\lambda_7}D^{-1}\nonumber\\
&=&D e^{i\theta_{23}\lambda_7}e^{i\theta_{13}^{M^{(-)}}\lambda_5}
\left[ {\Delta E_{32}+A \over 2}
 \mbox{\rm diag} \left(1,0,1\right)- {B^{(-)} \over 2}
 \mbox{\rm diag} \left(1,0,-1 \right)\right]\nonumber\\
&{\ }&\times e^{-i\theta_{13}^{M^{(-)}}\lambda_5}
e^{-i\theta_{23}\lambda_7}D^{-1},
\label{eqn:sch}
\end{eqnarray}

\hglue -0.6cm
where the unit matrix $\mbox{\rm diag} (E_2,E_2,E_2)$ which
contributes only to the overall phase has been subtracted
from $M$,
$\Delta E_{ij}\equiv\Delta m^2_{ij}/2E$,
$E$ is the neutrino energy,
$A\equiv\sqrt{2} G_F N_e(x)$ stands for the matter effect \cite{msw} of the
Earth, $D\equiv \mbox{\rm diag}\left(e^{i\delta},1,1 \right)$ and
the standard parametrization \cite{pdg} has been used for the MNS matrix
(\ref{eqn:mns})
\begin{eqnarray}
U&=&\left(
\begin{array}{ccc}
c_{12}c_{13} & s_{12}c_{13} &  s_{13}e^{-i\delta}\nonumber\\
-s_{12}c_{23}-c_{12}s_{23}s_{13}e^{i\delta} & 
c_{12}c_{23}-s_{12}s_{23}s_{13}e^{i\delta} & s_{23}c_{13}\nonumber\\
s_{12}s_{23}-c_{12}c_{23}s_{13}e^{i\delta} & 
-c_{12}s_{23}-s_{12}c_{23}s_{13}e^{i\delta} & c_{23}c_{13}\nonumber\\
\end{array}\right)\nonumber\\
&=& D
e^{i\theta_{23}\lambda_7}
e^{i\theta_{13}\lambda_5}
D^{-1}
e^{i\theta_{12}\lambda_2},
\end{eqnarray}

\hglue -0.6cm
with the Gell-Mann matrices 
\begin{eqnarray}
\lambda_2=\left(
\begin{array}{ccc}
0&-i& 0\\
i&0&0\\
0&0&0
\end{array}\right),
~\lambda_5=\left(
\begin{array}{ccc}
0&0& -i\\
0&0&0\\
i&0&0
\end{array}\right),
~\lambda_7=\left(
\begin{array}{ccc}
0&0& 0\\
0&0&-i\\
0&i&0
\end{array}\right).
\end{eqnarray}

\hglue -0.6cm
\begin{figure}
\vglue -0.2cm\hglue -0.8cm 
\epsfig{file=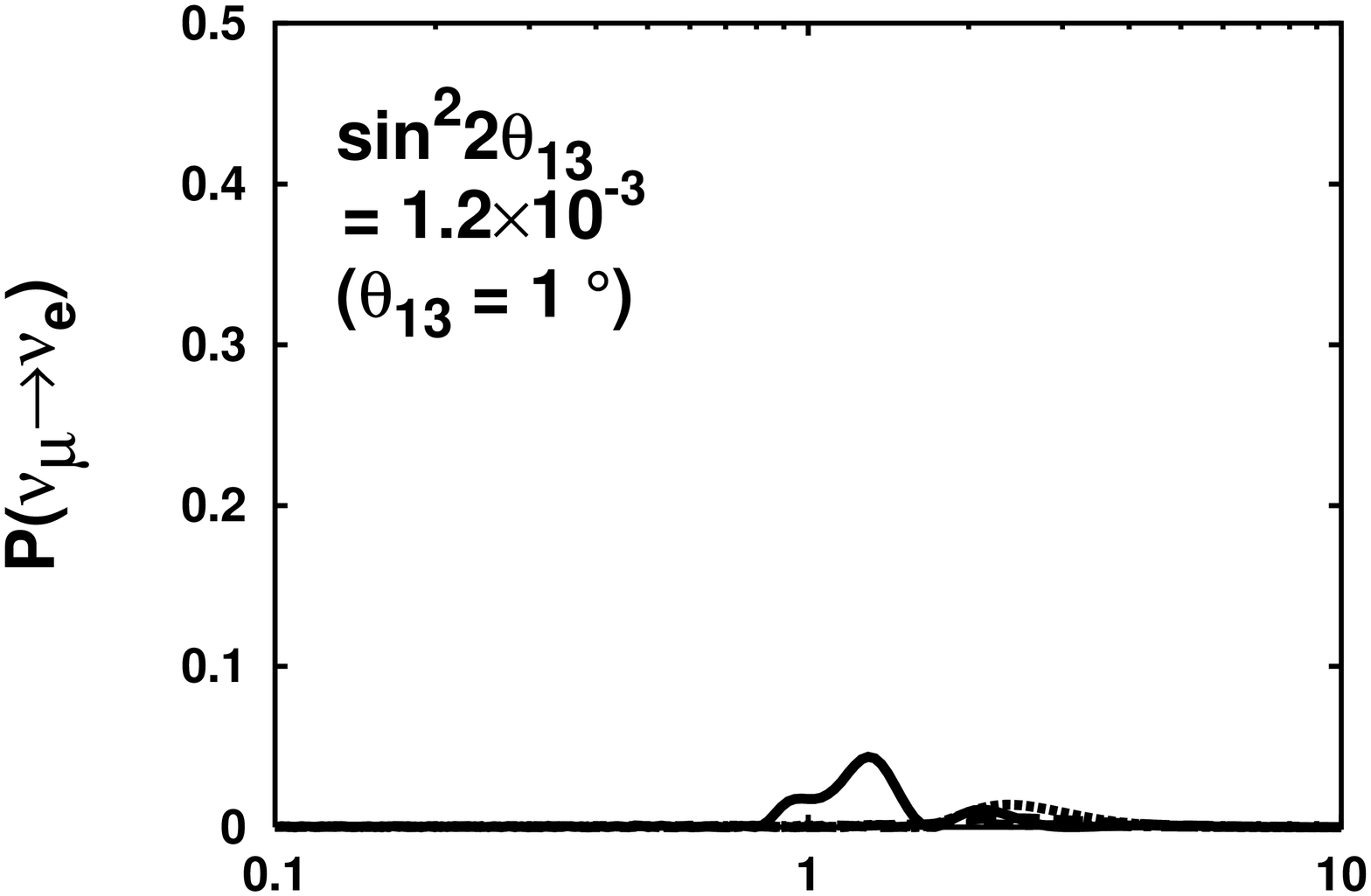,width=7.5cm}
\vglue -5.3cm\hglue 6.0cm \epsfig{file=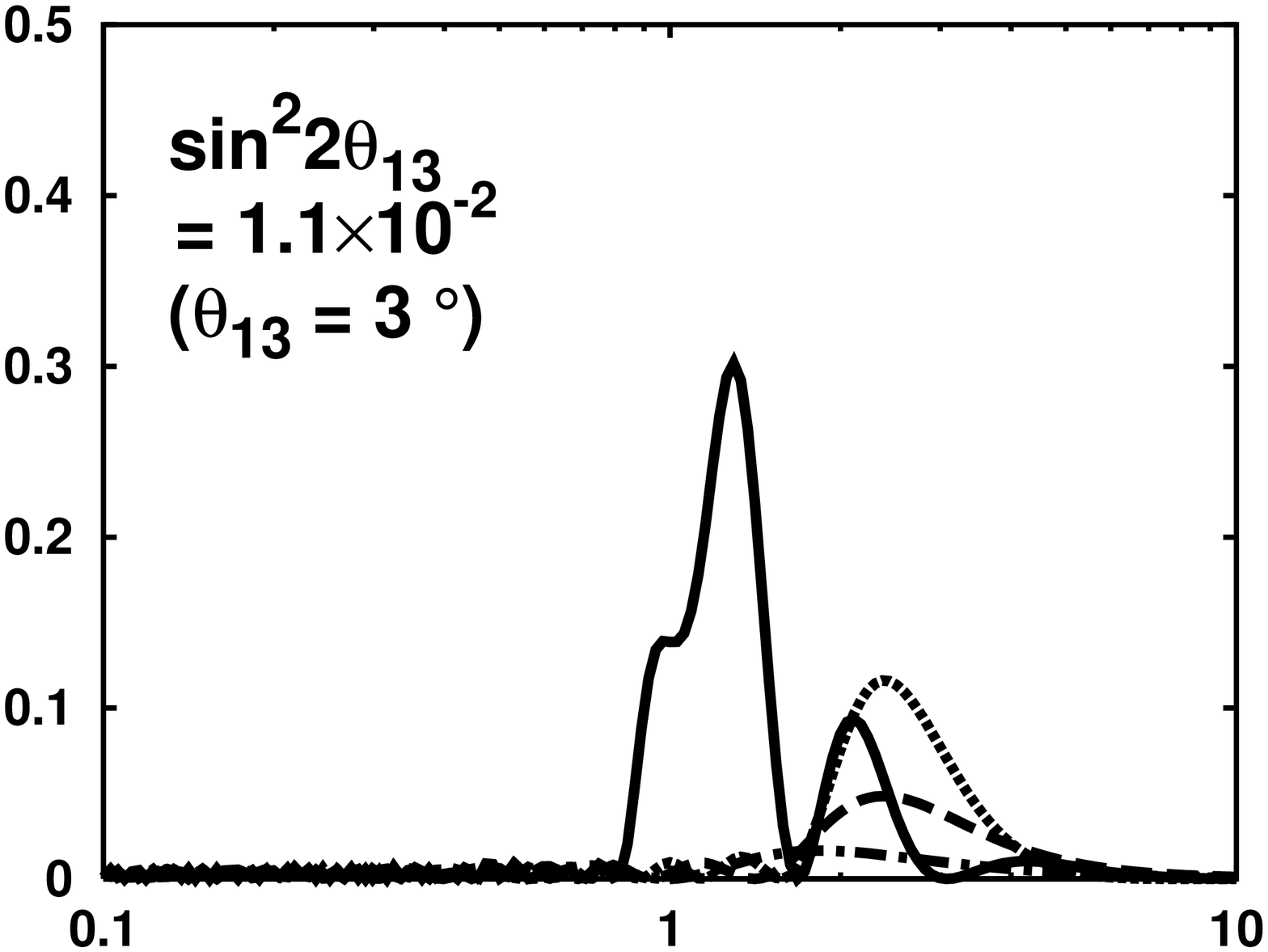,width=7.5cm}
\vglue -0.3cm\hglue -0.8cm 
\epsfig{file=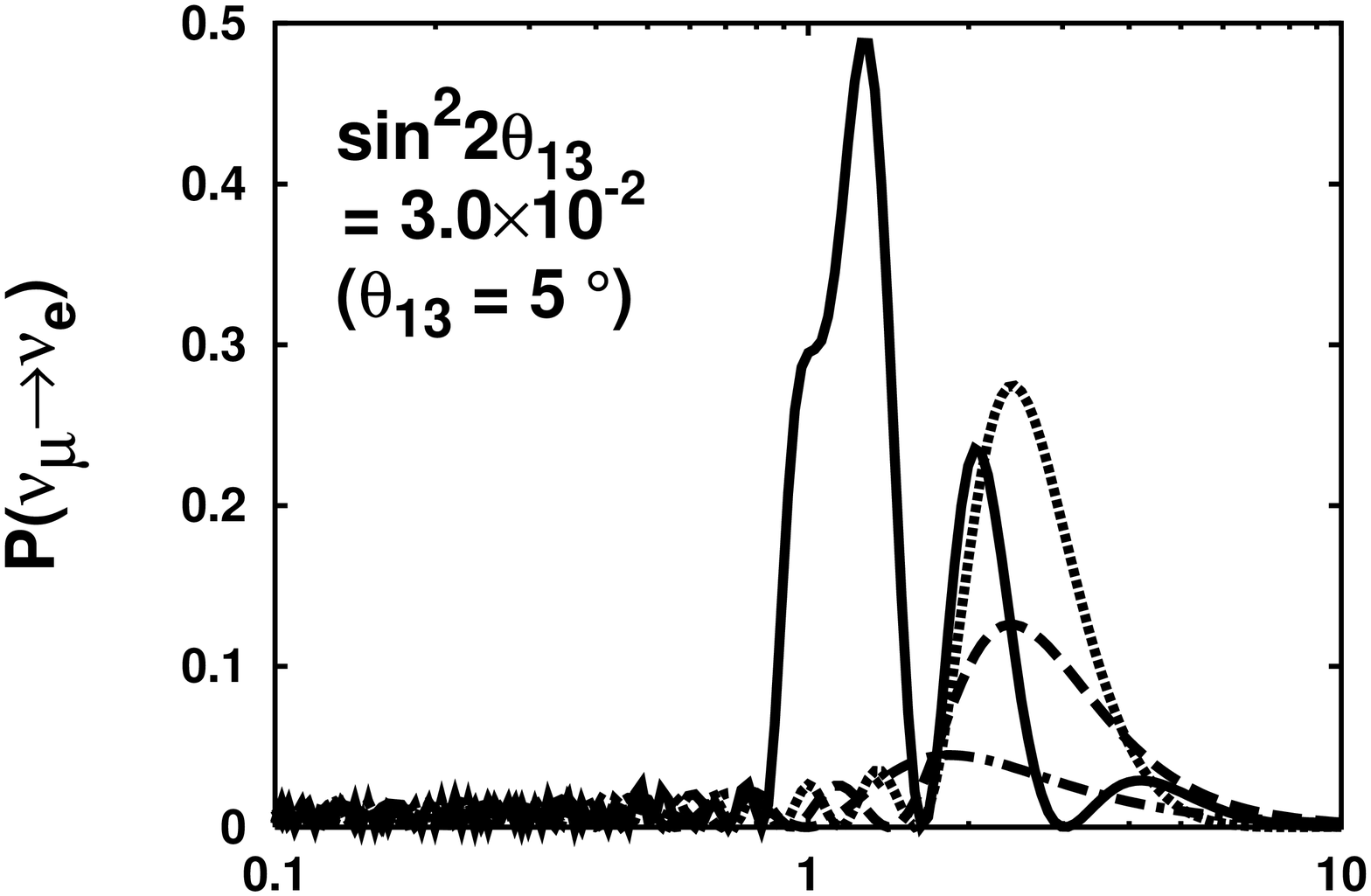,width=7.5cm}
\vglue -5.3cm\hglue 6.0cm\epsfig{file=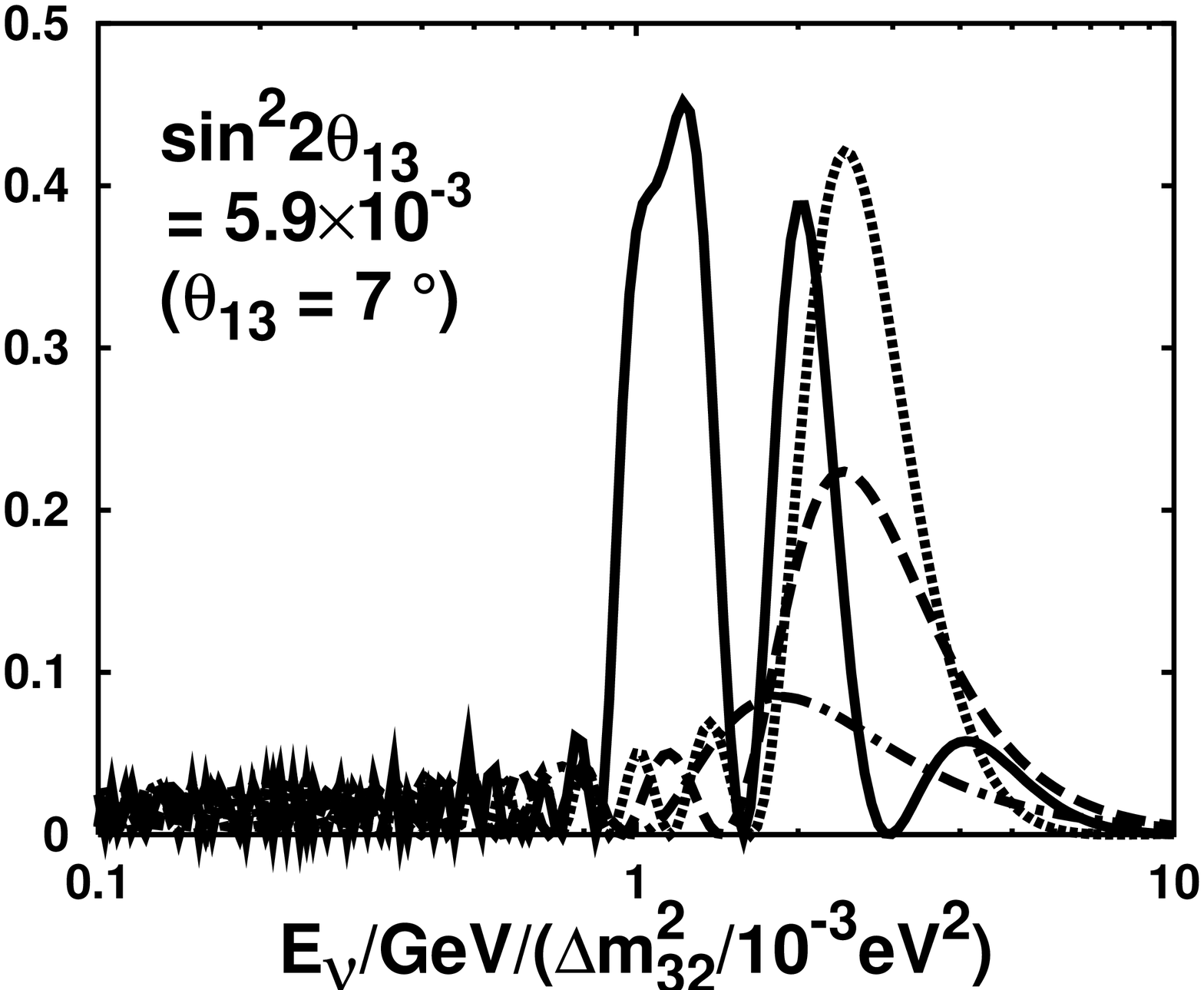,width=7.5cm}
\vglue -0.3cm\hglue -0.8cm 
\epsfig{file=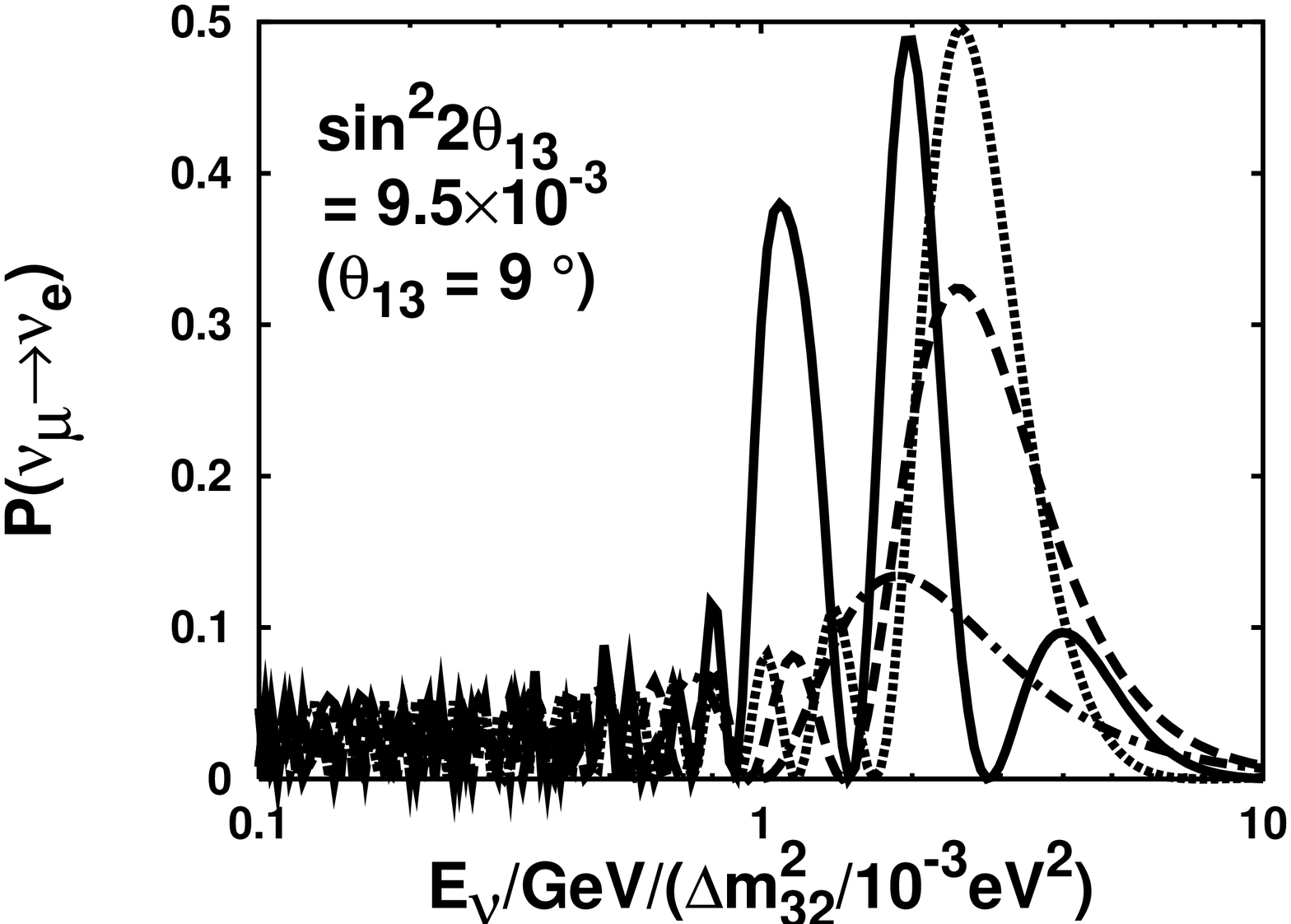,width=7.5cm}
\vglue -5.3cm\hglue 6.0cm\epsfig{file=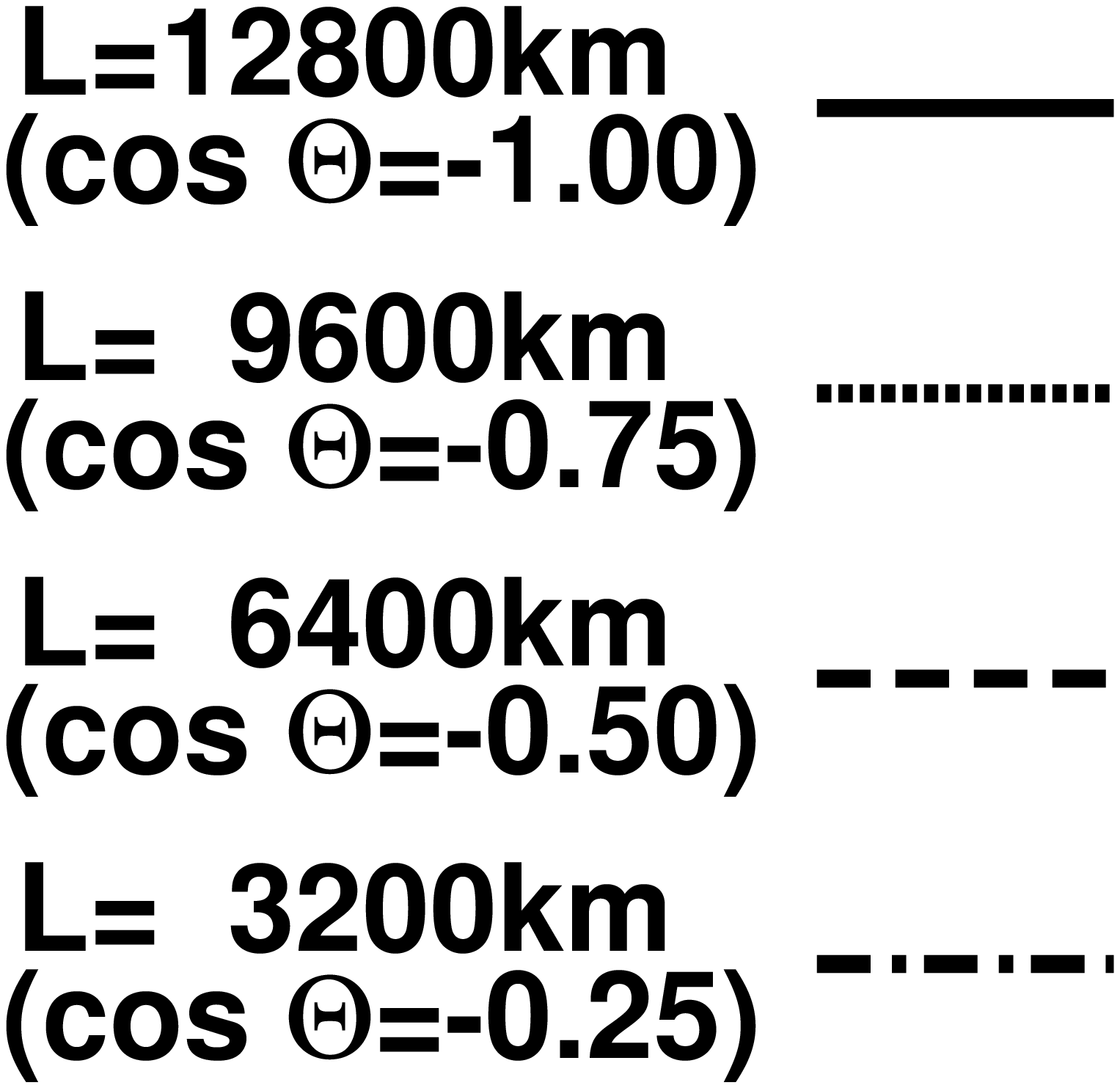,width=7.5cm}
\vglue 0.4cm
\caption{The appearance probability $P(\nu_\mu\rightarrow\nu_e)$
as a function of $E_\nu/\Delta m_{32}^2$ for $\Delta m_{32}^2>0$.
$\Delta m_{21}^2=0$, $\theta_{23}=\pi/4$ have been assumed.
In the case of $\Delta m_{32}^2<0$ one has to look at
$P({\bar \nu}_\mu\rightarrow{\bar \nu}_e)$ instead.}
\label{fig:mue}
\end{figure}

\hglue -0.6cm
$\theta_{13}^{M^{(\pm)}}$ is the mixing angle in matter given by
\begin{eqnarray}
\tan2\theta_{13}^{M^{(\pm)}}\equiv {\Delta E_{32}\sin2\theta_{13}
\over \Delta E_{32}\cos2\theta_{13}\pm A}
\label{eqn:thetam}
\end{eqnarray}

\hglue -0.6cm
as in the two flavor case \cite{msw} , and
\begin{eqnarray}
B^{(\pm)}\equiv \sqrt{\left(\Delta E_{32}\cos2\theta_{13}\pm A\right)^2
+\left(\Delta E_{32}\sin2\theta_{13}\right)^2},
\label{eqn:b}
\end{eqnarray}

\hglue -0.6cm
where $(+)$ sign is for antineutrinos,
as the sign of $A$ is reversed for antineutrinos.
Assuming the
constant density of the matter, the appearance probability
$P(\nu_\mu\rightarrow\nu_e)$ and
$P({\bar\nu}_\mu\rightarrow{\bar\nu}_e)$ in the three flavor framework
can be written as
\begin{eqnarray}
P(\nu_\mu\rightarrow\nu_e)=s^2_{23}\sin^22\theta_{13}^{M^{(-)}}
\sin^2\left({B^{(-)}L \over 2}\right)\nonumber\\
P({\bar\nu}_\mu\rightarrow{\bar\nu}_e)=s^2_{23}\sin^22\theta_{13}^{M^{(+)}}
\sin^2\left({B^{(+)}L \over 2}\right),
\label{eqn:mue}
\end{eqnarray}

\hglue -0.6cm
respectively.  Note that the only difference between the formulae
(\ref{eqn:mue}) and that in vacuum is $\sin^22\theta_{13}^{M^{(\pm)}}$
which is replaced by $\sin^22\theta_{13}$ in vacuum.
Thus the effective mixing angle
in matter is enhanced if $\Delta m_{32}^2>0$ and
$\Delta E_{32}\cos2\theta_{13}-A$
becomes small for some $E$.  On the other hand, if $\Delta m_{32}^2<0$,
then there is no enhancement in the probability
$P(\nu_\mu\rightarrow\nu_e)$, but the probability
$P({\bar \nu}_\mu\rightarrow{\bar \nu}_e)$ is enhanced instead.
The probability $P(\nu_e\rightarrow\nu_x)$ with matter effect of the
Earth has been discussed by many people in the framework of two
flavors \cite{earth2} and three flavors \cite{earth3}.  
The equation (\ref{eqn:mue})
is almost the same as that for two flavor case \cite{earth2},
the only difference being that the counterpart to $\nu_e$ is
the linear combination $s_{23}\nu_\mu+c_{23}\nu_\tau\simeq
(\nu_\mu+\nu_\tau)/\sqrt{2}$.

I have computed numerically the probability
$P(\nu_\mu\rightarrow\nu_e)$ for $\theta_{23}=\pi/4$,
$\theta_{13}=1^\circ,3^\circ,5^\circ,7^\circ,9^\circ$ (or
$\sin^22\theta_{13}=1.2\times10^{-3},1.1\times10^{-2},
3.0\times10^{-2},5.9\times10^{-2}, 9.5\times10^{-2}$) and for
$\cos\Theta=-1,-0.75,-0.5,-0.25$ (or $L$ = 12,800 km, 9,600 km, 6,400
km, 3,200 km) where the zenith angle $\Theta$, the neutrino path
length $L$ and the radius $R$ of the Earth are related by
$L=-2R\cos\Theta$.  The results are shown in Fig. \ref{fig:mue}.  The
shape of the probability $P(\nu_\mu\rightarrow\nu_e)$ is almost the
same as that for the two flavor oscillation \cite{earth2}, but it is
scaled by the normalization $s_{23}^2$ (cf. (\ref{eqn:mue})).  As can
be seen in Fig. \ref{fig:mue}, the case of $\cos\Theta=-1$
(or $L$ = 12,800 km) is most
advantageous to get the enhancement in the probability for
$\theta_{13}\gtrsim2^\circ$, while in the case of $\cos\Theta=-0.25$
it is difficult to see the enhancement for smaller values
of $\theta_{13}$.

To measure the probability in practical experiments, one has to
measure the momentum of the recoiled nucleon as well as that of the
outgoing charged lepton in a quasi elastic scattering
$\nu_\alpha+N\rightarrow\ell+N'$.  From Fig. \ref{fig:mue} we see that
the maximum probability is obtained for $E_\nu$/GeV$\simeq 1.2\times\Delta
m_{32}^2$/(10$^{-3}$eV$^2$) with $\cos\Theta=-1$ for each value of
$\theta_{13}$.  If we assume $\Delta
m_{32}^2\simeq$3.5$\times$10$^{-3}$eV$^2$ which is the best fit value
in the Superkamiokande atmospheric neutrino data \cite{SKatm}, then
the maximum probability is obtained for $E_\nu\simeq$ 4 GeV.  The
smaller $|\Delta m_{32}^2|$ becomes, the better it
works, since the cross section of quasi elastic
scatterings decreases as the neutrino energy increases \cite{ls}.

\section{A possible way to measure $\delta$}
There have been a lot of works which discussed CP violation in
neutrino oscillations \cite{cp1,cp2}.
From the oscillation probability in vacuum
\begin{eqnarray}
P(\nu_\alpha\rightarrow\nu_\beta;L)&=&\delta_{\alpha\beta}
-4\sum_{j<k}\mbox{\rm Re}\left(U_{\alpha j}U_{\beta j}^\ast
U_{\alpha k}^\ast U_{\beta k}\right)\sin^2\left(
{\Delta E_{jk}L \over 2}\right)\nonumber\\
&{\ }&+2\sum_{j<k}\mbox{\rm Im}\left(U_{\alpha j}U_{\beta j}^\ast
U_{\alpha k}^\ast U_{\beta k}\right)\sin\left(
\Delta E_{jk}L\right),
\label{eqn:prob}
\end{eqnarray}

\hglue -0.6cm
the CP violation in vacuum is given by
\begin{eqnarray}
&{\ }&P(\nu_\alpha\rightarrow\nu_\beta)
-P({\bar\nu}_\alpha\rightarrow{\bar\nu}_\beta)
\nonumber\\
&=&4\sum_{j<k}\mbox{\rm Im}\left(U_{\alpha j}U_{\beta j}^\ast
U_{\alpha k}^\ast U_{\beta k}\right)\sin\left(
\Delta E_{jk}L\right)\nonumber\\
&=&4~J
\left[\sin\left(\Delta E_{12}L\right)
+\sin\left(\Delta E_{23}L\right)
+\sin\left(\Delta E_{31}L\right)
\right],
\end{eqnarray}

\hglue -0.6cm
where
\begin{eqnarray}
J\equiv \mbox{\rm Im}\left(U_{\alpha 1}U_{\beta 1}^\ast
U_{\alpha 2}^\ast U_{\beta 2}\right)
\label{eqn:Jarlskog}
\end{eqnarray}

\hglue -0.6cm
is the Jarlskog factor, and
\begin{eqnarray}
\mbox{\rm Im}\left(U_{\alpha 1}U_{\beta 1}^\ast
U_{\alpha 2}^\ast U_{\beta 2}\right)=
\mbox{\rm Im}\left(U_{\alpha 2}U_{\beta 2}^\ast
U_{\alpha 3}^\ast U_{\beta 3}\right)=
\mbox{\rm Im}\left(U_{\alpha 3}U_{\beta 3}^\ast
U_{\alpha 1}^\ast U_{\beta 1}\right)
\end{eqnarray}

\hglue -0.6cm
has been used.
This Jarlskog factor, which is written as
\begin{eqnarray}
J=c_{13}\sin2\theta_{12}\sin2\theta_{13}\sin2\theta_{23}\sin\delta
\end{eqnarray}

\hglue -0.6cm
in the standard parametrization,
contains a small factor $\sin2\theta_{13}$ which
is constrained by the CHOOZ data ($\lesssim \sqrt{0.1}$), and a
factor $\sin2\theta_{12}$ which is small in the case of the small
mixing angle MSW solution ($\sin^22\theta_{12}\simeq 6\times
10^{-3}$).  So in general the Jarlskog factor is expected to be small.
In vacuum the CP violation happens to be the same as the T violation.

On the other hand, in the presence of matter, the expression
(\ref{eqn:prob}) for the probability is modified. 
The eigen matrix in matter
can be formally diagonalized by a unitary matrix $V$:
\begin{eqnarray}
U\,\mbox{\rm diag} \left(-\Delta E_{21},0,\Delta E_{32} \right) U^{-1}
+\mbox{\rm diag} \left(A,0,0 \right)
=V\,\mbox{\rm diag} \left(\zeta_1,\zeta_2,\zeta_3\right)V^{-1}.
\label{eqn:eigen}
\end{eqnarray}

\hglue -0.6cm
Assuming constant density, the oscillation probability can be written as
\begin{eqnarray}
P(\nu_\alpha\rightarrow\nu_\beta;L)&=&\delta_{\alpha\beta}
-4\sum_{j<k}\mbox{\rm Re}\left(V_{\alpha j}V_{\beta j}^\ast
V_{\alpha k}^\ast V_{\beta k}\right)\sin^2\left(
{\Delta \zeta_{jk}L \over 2}\right)\nonumber\\
&{\ }&+2\sum_{j<k}\mbox{\rm Im}\left(V_{\alpha j}V_{\beta j}^\ast
V_{\alpha k}^\ast V_{\beta k}\right)\sin\left(
\Delta \zeta_{jk}L\right),
\label{eqn:prob2}
\end{eqnarray}

\hglue -0.6cm
as in the case of the probability in vacuum.
The sign for the matter term $A$
is reversed for antineutrinos ${\bar\nu}_\alpha$ and the
unitary matrix $\bar V$ and the eigenvalues ${\bar \zeta}_j$
for ${\bar\nu}_\alpha$ are different from $V$ and $\zeta_j$ for
neutrinos $\nu_\alpha$.  Therefore
it is not illuminating to
see the CP violation in matter \cite{cp2}, since it
contains terms which vanish in the limit $\delta\rightarrow 0$
and those which do not:
\begin{eqnarray}
&{\ }&P(\nu_\alpha\rightarrow\nu_\beta)-
P({\bar\nu}_\alpha\rightarrow{\bar\nu}_\beta)\nonumber\\
&=&-4\sum_{j<k}\mbox{\rm Re}\left[V_{\alpha j}V_{\beta j}^\ast
V_{\alpha k}^\ast V_{\beta k}\sin^2\left(
{\Delta \zeta_{jk}L \over 2}\right)
-{\bar V}_{\alpha j}{\bar V}_{\beta j}^\ast
{\bar V}_{\alpha k}^\ast {\bar V}_{\beta k}\sin^2\left(
{\Delta {\bar \zeta}_{jk}L \over 2}\right)\right]\nonumber\\
&{\ }&+2\sum_{j<k}\mbox{\rm Im}\left[V_{\alpha j}V_{\beta j}^\ast
V_{\alpha k}^\ast V_{\beta k}\sin\left(
\Delta \zeta_{jk}L\right)
-{\bar V}_{\alpha j}{\bar V}_{\beta j}^\ast
{\bar V}_{\alpha k}^\ast {\bar V}_{\beta k}\sin\left(
\Delta {\bar \zeta}_{jk}L\right)\right],
\end{eqnarray}

\hglue -0.6cm
where $\Delta{\bar \zeta}_{jk}\equiv{\bar \zeta}_j-{\bar \zeta}_k$.
It has been pointed out \cite{tvio} that T violation
is useful to probe the CP violating phase in the presence of matter.
In fact from (\ref{eqn:prob2}) we have T violation in matter:
\begin{eqnarray}
\Delta P&\equiv&
P(\nu_\alpha\rightarrow\nu_\beta)- P(\nu_\beta\rightarrow\nu_\alpha)
\nonumber\\
&=&4\sum_{j<k}\mbox{\rm Im}\left(V_{\alpha j}V_{\beta j}^\ast
V_{\alpha k}^\ast V_{\beta k}\right)\sin\left(
\Delta \zeta_{jk}L\right)\nonumber\\
&=&4~{\cal J}
\left[\sin\left(\Delta \zeta_{12}L\right)
+\sin\left(\Delta \zeta_{23}L\right)
+\sin\left(\Delta \zeta_{31}L\right)
\right],
\end{eqnarray}

\hglue -0.6cm
where
\begin{eqnarray}
{\cal J}\equiv\mbox{\rm Im}\left(V_{\alpha 1}V_{\beta 1}^\ast
V_{\alpha 2}^\ast V_{\beta 2}\right)
\label{eqn:Jarlskogm}
\end{eqnarray}

\hglue -0.6cm
is the modified Jarlskog factor in matter and
$\Delta\zeta_{jk}\equiv\zeta_j-\zeta_k$.
Here I will consider T violation under two situations
where one of the mixing angles in (\ref{eqn:Jarlskogm}) is
enhanced due to the matter effect of the Earth in our scheme with
mass hierarchy.

\subsection{(a) $|\Delta E_{21}|\ll |A|\simeq|\Delta E_{32}|$}
First let us consider the case where
$|\Delta E_{21}|\ll |A|\simeq|\Delta E_{32}|$.
This case has been discussed by Arafune and Sato \cite{cp2}.
The unitary matrix which diagonalizes the eigen matrix
(\ref{eqn:eigen}) to first order in $|\Delta E_{21}/\Delta E_{32}|$
is given by
\begin{eqnarray}
V&\equiv& e^{i\theta_{23}\lambda_7}
e^{i\theta_{13}^{M^{(-)}}\lambda_5}
\exp\left\{i\Delta E_{21}\left[{1 \over u_-}{\tilde c}_{13}s_{12}c_{12}
(\lambda_1\sin\delta-\lambda_2\cos\delta)\right.\right.\nonumber\\
&{\ }&-\left.\left.{1 \over u_+-u_-}{\tilde s}_{13}{\tilde c}_{13}s_{12}^2\lambda_5
-{1 \over u_+}{\tilde s}_{13}s_{12}c_{12}
(\lambda_6\sin\delta+\lambda_7\cos\delta)
\right]\right\},
\end{eqnarray}

\hglue -0.6cm
where I have assumed $\Delta E_{32}>0$, $\theta_{13}^{M^{(-)}}$ and
$B^{(-)}$ are given in (\ref{eqn:thetam}) and (\ref{eqn:b}),
respectively, and
\begin{eqnarray}
\lambda_1&=&\left(
\begin{array}{ccc}
0&1& 0\\
1&0&0\\
0&0&0
\end{array}\right),
~\lambda_4=\left(
\begin{array}{ccc}
0&0&1\\
0&0&0\\
1&0&0
\end{array}\right),
~\lambda_6=\left(
\begin{array}{ccc}
0&0& 0\\ 0&0&1\\ 0&1&0\end{array}\right),\nonumber\\ u_\pm&\equiv&{1
\over 2}(\Delta E_{32}+A\pm B^{(-)}),\nonumber\\
{\tilde \theta}_{13}&\equiv&\theta_{13}-\theta_{13}^{M^{(-)}},
~{\tilde s}_{13}\equiv\sin{\tilde \theta}_{13},
~{\tilde c}_{13}\equiv\cos{\tilde \theta}_{13}.
\end{eqnarray}

\hglue -0.6cm
The modified Jarlskog factor ${\cal J}_1$ in this case is given by
\begin{eqnarray}
{\cal J}_1&=&{2\Delta E_{21} \over A\cos2\theta_{13}}
{c_{13} \over 8}\sin2\theta_{12}\sin2\theta_{13}^{M^{(-)}}\sin2\theta_{23}
\sin\delta.
\label{eqn:j1}
\end{eqnarray}

\hglue -0.6cm
If $\Delta E_{32}<0$ then we have to look at $\Delta \bar{P}\equiv
P(\bar{\nu}_\alpha\rightarrow\bar{\nu}_\beta)-
P(\bar{\nu}_\beta\rightarrow\bar{\nu}_\alpha)$ instead, and
$\theta_{13}^{M^{(-)}}$ and $B^{(-)}$ have to be replaced by
$\theta_{13}^{M^{(+)}}$ and $B^{(+)}$, respectively.  ${\cal J}_1$ is
similar to the Jarlskog factor $J$ in vacuum, but one important
difference between $J$ and ${\cal J}_1$ is that ${\cal J}_1$ has
$\sin2\theta_{13}^{M^{(-)}}$ which could be enhanced by the matter
effect of the Earth.  Another difference is that ${\cal J}_1$ has a
factor $\Delta E_{21}/A$ whose absolute value is small by assumption
and therefore the T violating effect under this condition (a) is
supposed to be small.

\subsection{(b) $|\Delta E_{21}|\simeq |A|\ll|\Delta E_{32}|$}
Next let us consider the case where
$|\Delta E_{21}|\simeq |A|\ll|\Delta E_{32}|$.
In this case it is more convenient to adopt the original
parametrization for $U$ proposed by Kobayashi and Maskawa \cite{km}:
\begin{eqnarray}
U&=&\left(
\begin{array}{ccc}
c_1 & -s_1c_3 &  -s_1s_3\nonumber\\
s_1c_2 & 
c_1c_2c_3-s_2s_3e^{i\delta} &
c_1c_2s_3+s_2c_3e^{i\delta}\nonumber\\
s_1s_2 & 
c_1s_2c_3+c_2s_3e^{i\delta} &
c_1s_2s_3-c_2c_3e^{i\delta}\nonumber\\
\end{array}\right)\nonumber\\
&=& 
e^{-i\theta_2\lambda_7}
e^{-i\theta_1\lambda_2}
{\cal D}
e^{i\theta_3\lambda_7},
\end{eqnarray}

\hglue -0.6cm
where ${\cal D}\equiv\mbox{\rm diag} \left(1,1,-e^{i\delta}\right)$.
The eigen matrix (\ref{eqn:eigen}) is decomposed as the zero-th order
term $M_0$ and the first order contribution $M_1$ with respect to
$|\Delta E_{21}/\Delta E_{32}|$:
\begin{eqnarray}
U \mbox{\rm diag} \left(-\Delta E_{21},0, \Delta E_{32}\right) U^{-1}
+\mbox{\rm diag} \left(A,0,0 \right)
=U(M_0+M_1)U^{-1}
\end{eqnarray}

\hglue -0.6cm
with
\begin{eqnarray}
M_0&\equiv&(0,0,\Delta E_{32}),\nonumber\\
M_1 &\equiv& \mbox{\rm diag} \left(-\Delta E_{21},0,0\right)
+U^{-1}\mbox{\rm diag} \left(A,0,0 \right)U\nonumber\\
&=&U^{-1}e^{-i\theta_2\lambda_7}
e^{-i\theta_1^M\lambda_2}
\mbox{\rm diag} \left(t_-,t_+,0 \right)
e^{i\theta_1^M\lambda_2}e^{i\theta_2\lambda_7}U\nonumber\\
&=&e^{-i\theta_3\lambda_7}{\cal D}^{-1}e^{i{{\tilde\theta}_1}\lambda_2}
\mbox{\rm diag} \left(t_-,t_+,0 \right)e^{-i{{\tilde\theta}_1}\lambda_2}
{\cal D}e^{i\theta_3\lambda_7}\nonumber\\
&\equiv&\mbox{\rm diag} \left(\xi_1,\xi_2,\xi_3\right)
+\eta_1\lambda_1+\eta_4\lambda_4+\eta_6\lambda_6,
\end{eqnarray}

\hglue -0.6cm
where
\begin{eqnarray}
t_\pm&\equiv&{1 \over 2}
\left(A-\Delta E_{21}\pm
\sqrt{(\Delta E_{21}\cos2\theta_1-A)^2+(\Delta E_{21}\sin2\theta_1)^2}
\right),\nonumber\\
&{\ }&\tan2\theta_1^M\equiv {\Delta E_{21}\sin2\theta_1
\over \Delta E_{21}\cos2\theta_1- A},\nonumber\\
\eta_1&\equiv&(t_+-t_-){\tilde c}_1{\tilde s}_1c_3,~~
\eta_4\equiv(t_+-t_-){\tilde c}_1{\tilde s}_1s_3,~~
\eta_6\equiv(t_-{\tilde s}_1^2+t_+{\tilde c}_1^2)c_3s_3,\nonumber\\
\xi_1&\equiv&t_-{\tilde c}_1^2+t_+{\tilde s}_1^2,~~
\xi_2\equiv(t_-{\tilde s}_1^2+t_+{\tilde c}_1^2)c_3^2,~~
\xi_3\equiv(t_-{\tilde s}_1^2+t_+{\tilde c}_1^2)s_3^2,\nonumber\\
{\tilde\theta}_1&\equiv&\theta_1-\theta_1^M,~~
{\tilde s}_1\equiv\sin{\tilde\theta}_1,~~
{\tilde c}_1\equiv\cos{\tilde\theta}_1.
\label{eqn:phi}
\end{eqnarray}

\hglue -0.6cm
Using perturbation theory in $|\Delta E_{21}/\Delta E_{32}|$, it can
be shown that the unitary matrix
\begin{eqnarray}
V\equiv U~e^{i(\eta_4\lambda_5+\eta_6\lambda_7)/\Delta E_{32}}
e^{-i\psi_1^M\lambda_2}
\label{eqn:v2}
\end{eqnarray}

\hglue -0.6cm
diagonalizes
$U(M_0+M_1)U^{-1}$ to first order in $|\Delta E_{21}/\Delta E_{32}|$:
\begin{eqnarray}
U\left(M_0+M_1\right)U^{-1}=V~\mbox{\rm diag} \left({\xi_1+\xi_2 \over 2}+b,
{\xi_1+\xi_2 \over 2}-b,\xi_3+\Delta E_{32}
\right)V^{-1},\nonumber\\
\end{eqnarray}

\hglue -0.6cm
where
\begin{eqnarray}
b&\equiv&\sqrt{\left({\xi_1-\xi_2 \over 2}\right)^2+\eta_1^2}\nonumber\\
\tan2\psi_1^M&\equiv& {2\eta_1 \over \xi_1-\xi_2}
={2(t_+-t_-){\tilde c}_1{\tilde s}_1c_3 \over
t_-{\tilde c}_1^2+t_+{\tilde s}_1^2-
(t_-{\tilde s}_1^2+t_+{\tilde c}_1^2)c_3^2}.
\label{eqn:psi}
\end{eqnarray}

\hglue -0.6cm
In the limit $|\Delta E_{21}/\Delta E_{32}|\rightarrow 0$,
(\ref{eqn:v2}) becomes
\begin{eqnarray}
V&\simeq& U~e^{-i\psi_1^M\lambda_2}\nonumber\\
&=&\left(
\begin{array}{ccc}
c_\psi U_{e1}+s_\psi U_{e2}&-s_\psi U_{e1}+c_\psi U_{e2} &  U_{e3}\\
c_\psi U_{\mu 1}+s_\psi U_{\mu 2}&-s_\psi U_{\mu 1}+c_\psi U_{\mu 2} &
U_{\mu 3} \\
c_\psi U_{\tau 1}+s_\psi U_{\tau 2}&-s_\psi U_{\tau 1}+c_\psi U_{\tau 2} &
U_{\tau 3}
\end{array}\right),
\end{eqnarray}

\hglue -0.6cm
where $s_\psi\equiv\sin\psi_1^M,~~c_\psi\equiv\cos\psi_1^M$.
Notice that $\psi_1^M$ survives even in the limit
$|\Delta E_{21}/\Delta E_{32}|\rightarrow 0$.
Therefore the modified Jarlskog factor ${\cal J}_2$ is given by
\begin{eqnarray}
{\cal J}_2&\simeq&\mbox{\rm Im}\left[
(c_\psi U_{e1}+s_\psi U_{e2})
(-s_\psi U_{e1}+c_\psi U_{e2})\right.\nonumber\\
&{\ }&\times
\left.
(c_\psi U_{\mu 1}+s_\psi U_{\mu 2})^\ast
(-s_\psi U_{\mu 1}+c_\psi U_{\mu 2})\right]\nonumber\\
&=&{1 \over 2}\left[{1 \over 2}\sin2\psi_1^M
(1-U_{e3}^2)-U_{e1}U_{e2}\cos2\psi_1^M\right]
U_{e3}\sin2\theta_2\sin\delta.
\end{eqnarray}

\hglue -0.6cm
In the case of the small mixing angle MSW solution for which
we may neglect small factors $U_{e3}^2$ and $U_{e2}$, we have
\begin{eqnarray}
{\cal J}_2&=&{1 \over 4}\sin2\psi_1^M\sin2\theta_2U_{e3}\sin\delta
\simeq{1 \over 4}\sin2\theta_1^M\sin2\theta_2U_{e3}\sin\delta,
\label{eqn:j2}
\end{eqnarray}

\hglue -0.6cm
where we have used the fact $\tan2\psi_1^M\simeq\tan2\theta_1^M$ which
follows from (\ref{eqn:psi}) when $|\theta_3|\ll 1$.  (\ref{eqn:j2})
could be large since ${\cal J}_2$ does not contain the suppression
factor $|\Delta E_{21}/\Delta E_{32}|$ like in ${\cal J}_1$,
$\sin2\theta_1^M$ could be enhanced by the matter effect of the Earth
(cf. (\ref{eqn:phi})), and
the only factor which is always small is $U_{e3}$
($|U_{e3}|\lesssim \sqrt{0.1}/2$).
Note that we know $\Delta E_{21}>0$ from the solar neutrino deficit.

\subsection{Numerical analysis}
\begin{figure}
\vglue -3.9cm
\hglue 2cm 
\epsfig{file=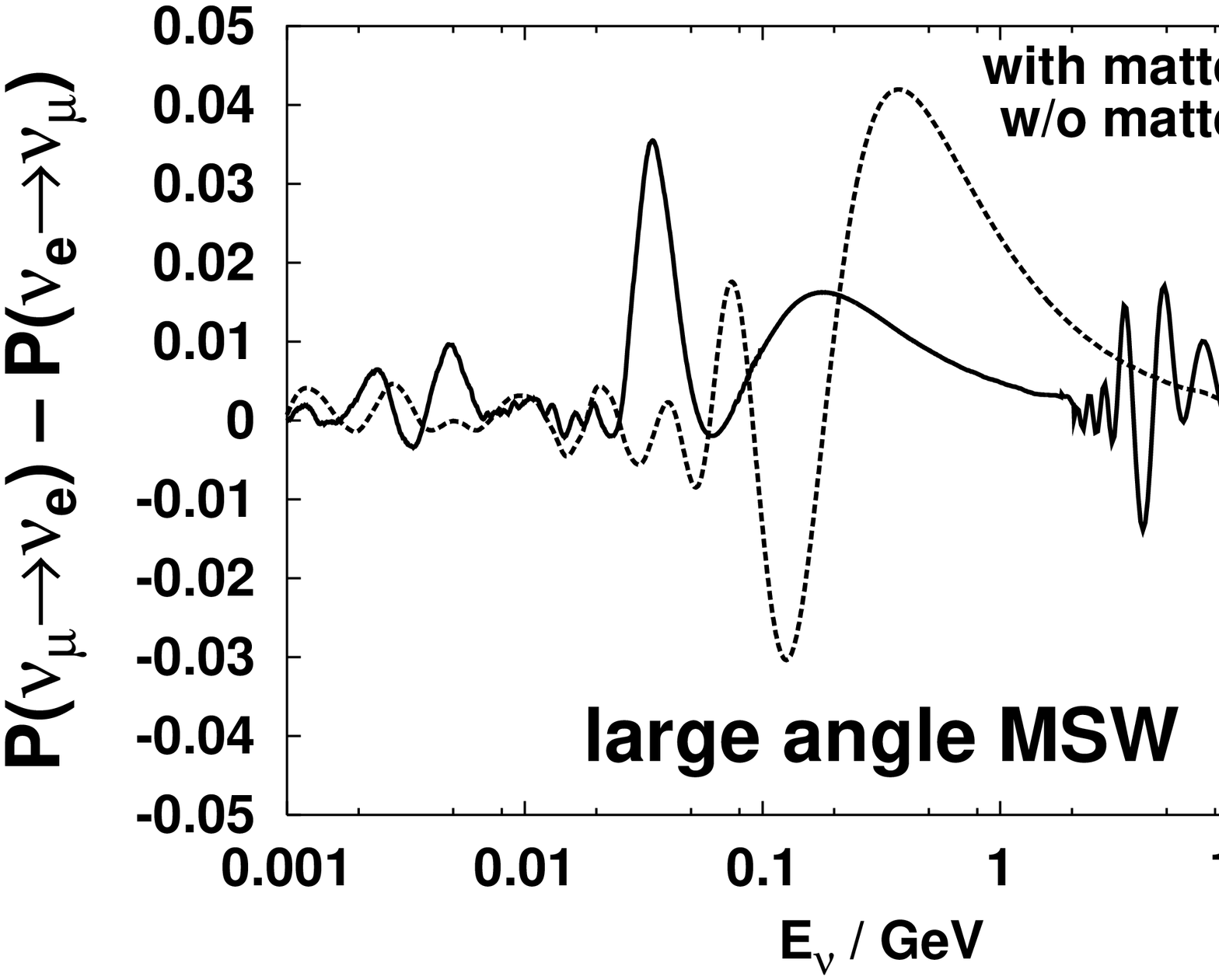,width=6cm}
\vglue -3.2cm
\hglue 2cm 
\epsfig{file=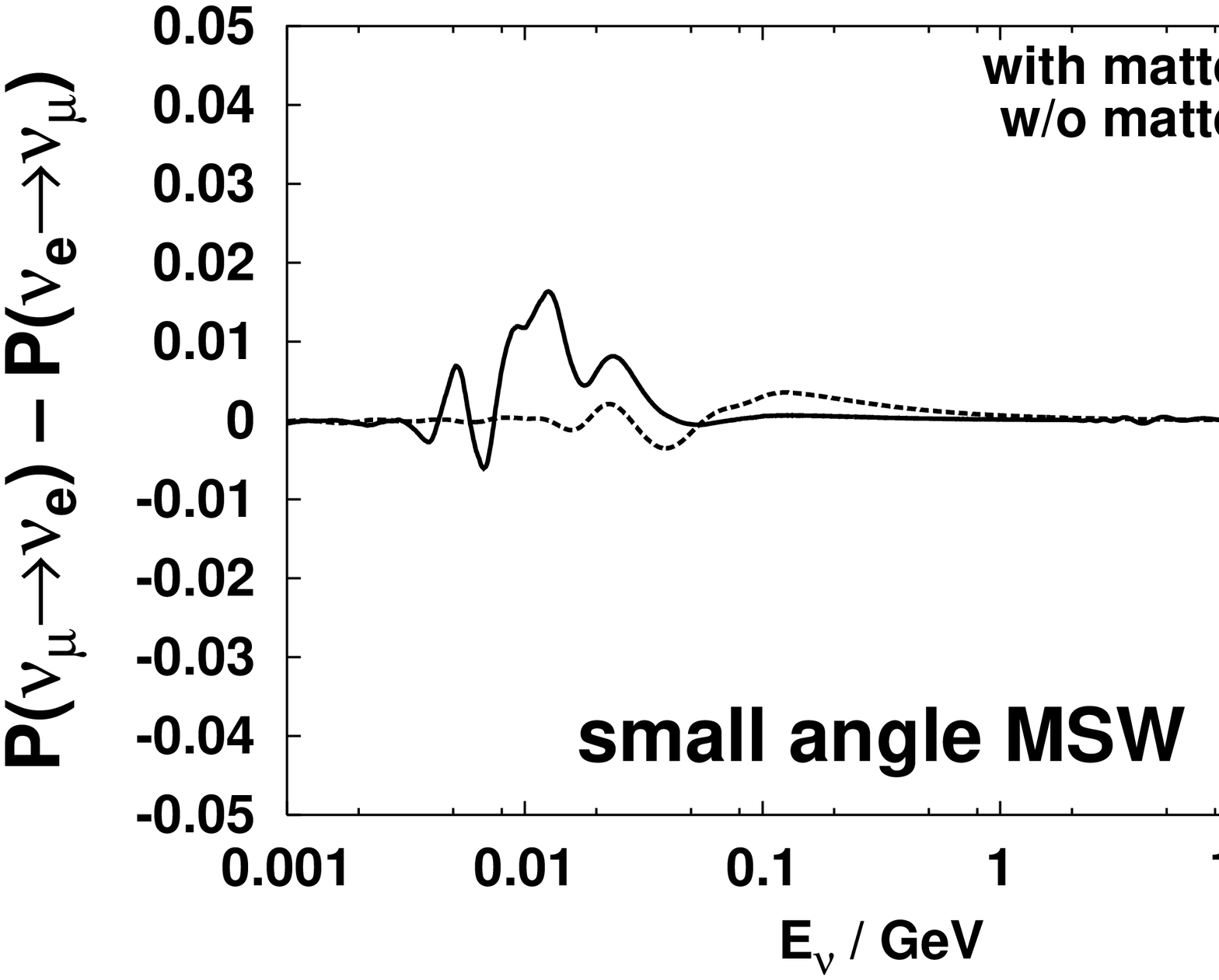,width=6cm}
\caption{ T violation
$P(\nu_\mu\rightarrow\nu_e)-P(\nu_e\rightarrow\nu_\mu)$ in matter
(solid lines) and in vacuum (dashed lines) for the large ($\Delta
m_{21}^2=1.8\times 10^{-5}$eV$^2$, $\sin^22\theta_{12}=0.76$) and
small ($\Delta m_{21}^2=5.4\times 10^{-6}$eV$^2$,
$\sin^22\theta_{12}=0.006$) angle MSW solutions.
$\sin^22\theta_{13}$ = 0.036 and the set of the best
fit parameters $(\Delta m^2_{32},\sin^22\theta_{23})$ = ($3.5\times
10^{-3}$eV$^2$, 1.0) for the atmospheric neutrino data
have been taken as reference values, maximum T violation
$\delta=\pi/2$ has been assumed and rapid oscillations due to larger
eigenvalues have been averaged over.}
\label{fig:t}
\end{figure}
The results by numerical calculations are given in Fig. 5 where rapid
oscillations coming from the larger eigenvalues
are averaged over, maximal T violation is assumed ($\delta=\pi/2$) and
the best fit values \cite{bks} of the large and small angle MSW
solutions have been chosen for the set of parameters $(\Delta
m^2_{21},\sin^22\theta_{12})$, respectively.  In the case of the large
mixing angle MSW solution there is some resonance for $E_\nu\sim$
several GeV and $E_\nu\sim$ $\cal O$(0.1) GeV which satisfy the
conditions (a) and (b), respectively.  In the case of the small mixing
angle MSW solution, the enhancement is hardly visible for the energy
$E_\nu\sim$ several GeV which satisfies the condition (a) but it is
significant for $E_\nu\sim$ $\cal O$(10) MeV which satisfies the condition (b).
We see from Fig. \ref{fig:t} that T violating effects are significant
for relatively low neutrino energy, i.e., $E_\nu\sim$ $\cal O$(0.1)
GeV and $E_\nu\sim$ $\cal O$(10) MeV in the case of the large and
small angle MSW solutions, respectively.  For the former case, the
enhancement around $E_\nu\sim$ several GeV
might be observable at neutrino factories \cite{nf}, where intense
beams of $\nu_e$ and $\nu_\mu$ (or $\bar\nu_e$ and $\bar\nu_\mu$) are
produced.  For the latter case, it would be very difficult to see
these effects unless we can produce intense
beams of neutrinos with energy $E_\nu\sim$
$\cal O$(10) MeV.  It should be also mentioned that it is hopeless to
see T violation effects in the case of the vacuum oscillation
solutions, since $\Delta m^2_{21}$ is extremely small.

\section{Conclusions}
To summarize, I have shown that it is possible to probe
the small value of $\theta_{13}$ down to $\sin^22\theta_{13}\gtrsim$0.01
in very long baseline
experiments by looking for $\nu_e$ (in the case of $\Delta m_{32}^2>0$)
or ${\bar \nu}_e$ (in the case of $\Delta m_{32}^2<0$)
appearance which is enhanced due to the matter effect of the Earth.
If we can produce very intense beams of $\nu_\mu$ and $\nu_e$
with low energy then we may be able to determine the CP violating
phase $\delta$ by looking at T violating effects in the case of
the MSW solutions.

\section{Acknowledgement}
The author would like to thank Prof. H. Czyz and
other organizers for invitation and hospitality
during the conference.
He would also like to thank Summer Institute
99 at Yamanashi, Japan for hospitality during part of this work
and the participants of SI99 for discussions.
This work was supported in part by a
Grant-in-Aid for Scientific Research of the Ministry of Education,
Science and Culture, \#09045036, \#10640280.

\end{document}